\documentclass[pdflatex,sn-mathphys-ay]{sn-jnl}

\usepackage{graphicx}%
\usepackage{multirow}%
\usepackage{amsmath,amssymb,amsfonts}%
\usepackage{amsthm}%
\usepackage{mathrsfs}%
\usepackage[title]{appendix}%
\usepackage{xcolor}%
\usepackage{textcomp}%
\usepackage{manyfoot}%
\usepackage{booktabs}%
\usepackage{algorithm}%
\usepackage{algorithmicx}%
\usepackage{algpseudocode}%
\usepackage{listings}%
\usepackage{comment}

\usepackage{bm}

\theoremstyle{thmstyleone}%
\theoremstyle{thmstyletwo}%

\theoremstyle{thmstylethree}%

\begin{document}

\title[Article Title]{Calibrating simplified vine copulas with
a noise contrastive estimation approach}


\author[1]{\fnm{Michael Denis } \sur{Kraus}}\email{michael.d.kraus@tum.de}

\author[2]{\fnm{David} \sur{Huk}}\email{David.Huk@warwick.ac.uk}

\author*[1]{\fnm{Claudia} \sur{Czado}}\email{cczado@ma.tum.de}

\affil*[1]{\orgdiv{Department of Mathematics, School of Computation, Information and Technology}, \orgname{Technical University of Munich}, \orgaddress{\street{Boltzmannstr. 3}, \city{Garching}, \postcode{85748}, \country{Germany}}}

\affil[2]{\orgdiv{Department of Statistics}, \orgname{University of Warwick}, \orgaddress{\city{Coventry}, \postcode{CV4 7AL}, \country{United Kingdom}}}


\abstract{Vine copulas provide a flexible framework for modeling complex multivariate dependence structures using only bivariate building blocks. Their practical success relies heavily on the simplifying assumption, which restricts conditional pair copulas to be independent of the specific conditioning values. While this assumption greatly facilitates estimation, it may lead to model misspecification in applications with pronounced varying conditional dependence. 
We propose a novel calibration strategy for simplified vine copula models based on observation-specific correction factors. These factors are derived using noise contrastive estimation (NCE), a supervised learning technique for density estimation that reframes the problem as a binary classification task with an easily sampled noise distribution. Treating the fitted simplified vine copula as the noise model, the NCE approach yields corrected log-likelihood estimates for individual observations, thereby locally adjusting the simplified vine toward the underlying data-generating dependence structure.
Simulation studies demonstrate that the proposed calibration provides sensible and effective adjustments, improving model accuracy when the simplifying assumption is violated while remaining neutral when the simplified model is adequate. Two real-data applications further illustrate the practical benefits of the method. The results highlight NCE-based calibration as a promising tool to enhance simplified vine copula models without abandoning their computational tractability.}

\keywords{dependence, vine copulas, simplifying assumption, noise contrastive estimation, supervised learning}



\maketitle

\section{Introduction}\label{sec1}
\label{sec:intro}

Vine copula based modeling has become a very popular tool in multivariate statistics to model complex dependency patterns. It follows the copula approach first proposed by Sklar (1959) and constructs a multivariate copula by conditioning and using only bivariate copulas also called pair copulas as building blocks. These represent conditional and unconditional bivariate dependencies. To make the stepwise estimation procedure proposed in \cite{aas2009pair} computationally tractable, a simplifying assumption is utilized. It assumes that conditional pair copulas depend on the conditioning values only through their arguments, but the copula itself remains independent of the conditioning values. This restriction is somewhat mitigated by allowing a multitude of parametric bivariate copula families (\cite{Joe}), as well as non-parametric bivariate copulas. Early approaches to allow the conditional pair copulas to depend on the conditioning values allowed the copula parameter to depend on it. (\cite{vatter2018generalized}, \cite{acar2019flexible}). For the current state of discussion see \cite{nagler2025simplified} and the references therein. Here, we propose a new approach to calibrate a fitted simplified vine copula model via an observation-specific correction factor. To determine such a correction factor we use  the noise contrastive estimation approach of \cite{Gutmann2012} developed in machine learning. It is a supervised learning method  used to learn unknown multivariate densities by reframing the problem as a binary classification task. It requires a noise model from which one can easily sample. The approach gives posterior classification probability estimates which are utilized to estimate the unknown log-likelihood of each observation. In our case the noise distribution is the fitted simplified vine distribution, which is then corrected for each observation.  We show that this approach gives sensible observation-specific adjustments in simulations. We illustrate their use in two real data applications. One shows that no calibration is needed, while the other shows that the adjustments provide a major improvement.

The paper is organized as follows: After the introduction, the necessary background on vine copulas and the simplifying assumption is provided in Section 2. In Section 3, the noise contrastive estimation approach is discussed, while the proposed calibration is given in Section 4. An illustration and simulation study is contained in Section 5. The two data applications are given in Section 6. The paper closes with a short conclusion and outlook section.

\section{Vine copulas and the simplifying assumption}
\label{sec:vine}
Many applications need to account for non-linear relationships and extreme co-movements in their statistical modeling. One way to include these is to follow a copula approach \citep{Sklar1959}.  This approach allows to separate the marginal behavior from the dependence structure. More precisely,
a $d$-dimensional copula $C$ is a multivariate
distribution function on  $[0,1]^d$ with uniformly distributed marginals and corresponding density $c$ in the absolutely continuous case, assumed from now on. In this case the copula is unique and  the
joint distribution function $F$ of the random vector $\bm X=(X_1,\ldots, X_d)$ with 
marginal distribution functions $F_j, j=1,\ldots,d$,
can be expressed as
\begin{align}
\label{sklar-cdf}
F(x_1,...,x_d) &= C(F_1(x_1),...,F_d(x_d)).
\end{align}

The joint
density is
$
f(x_1,...,x_d) = c(F_1(x_1),...,F_d(x_d)) f_1(x_1)...f_d(x_d),
$
where $f_j$ is the marginal density of $X_j$. Hence, any copula can be linked with arbitrary margins to define a distribution for $\bm X$. By inverting \eqref{sklar-cdf}, any $F$ can be used to define a copula $C$, giving rise to the Gaussian and the Student $t$ copula. Further, generator functions can be used to directly define a copula, such as the class of Archimedean copulas with Gumbel, Clayton, and Frank copulas as special cases (\cite{Nelsen,Joe}).
For pairs of variables, dependence and tail behavior are measured by 
Kendall's $\tau$ and upper and lower tail dependence coefficients. For $d=2$, the Gaussian or Frank copula have zero tail dependence coefficients, while the Student $t$ copula has symmetric tail dependence coefficients. The Clayton (Gumbel) copula has lower (upper) tail dependence (see  for example Table 3.3 of \cite{Czado}).

For small numbers of parameters, joint maximum likelihood is used to estimate the marginal and copula parameters in \eqref{sklar-cdf}. However, a two-step approach is often followed: First, estimate parametrically or non-parametrically - either by the empirical or kernel-based distribution - the marginal distributions functions $\hat{F}_j, j=1,\ldots,d$, separately using i.i.d. observations $\bm x_i=(x_{i1},\ldots, x_{id})^\top$ for $i=1,\ldots, n$. Second, utilizing the estimated distribution functions $\hat{F}_j$, create pseudo copula data by setting
\begin{equation}
\label{eq:pseudo}
\bm u_i=(u_{i1},\ldots,u_{id})^\top:=(\hat{F}_1(x_{i1}),\ldots,\hat{F}_d(x_{id}))^\top \mbox{ for } i=1,...,n.
\end{equation}
 Copula parameters are then estimated using the pseudo copula data in \eqref{eq:pseudo}. 

 While there are many bivariate parametric copula families, this does not hold for $d>2$. This led \cite{Joe96} to apply conditioning to construct multivariate copulas using only bivariate (pair) copulas as building blocks giving a pair copula construction (PCC) in terms of distribution
functions. \cite{BedfordCooke2001} 
developed constructions in terms of densities and defined a general vine tree structure to identify all possible constructions.
A vine tree structure is a sequence of $d$ linked trees $T_i$ with node set $N_i$ and edge set $E_i$ for $i=1,\ldots,d$. Each edge $e \in E_i$ corresponds to a bivariate copula $C$, modeling the dependence between variables $X_{a_e}, X_{b_e}$, given a conditioning set $X_{D_e}$. 
More details can be found in Chapter 5 of \cite{Czado}.

\begin{figure}[ht]
		\centering
		\includegraphics[width=0.6\textwidth ]{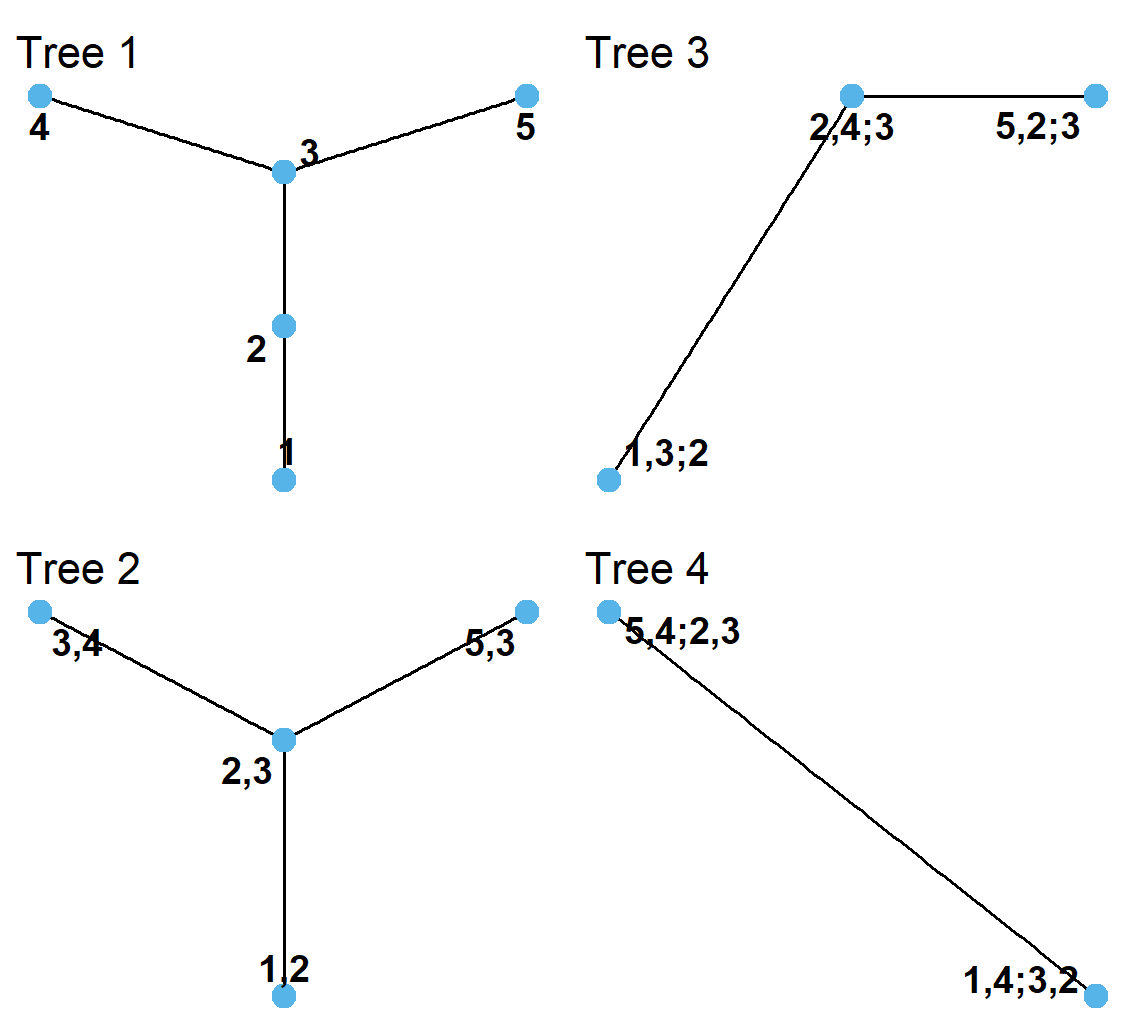}
		\caption{Example of a vine tree structure in five dimensions.}
		\label{fig:introTreePlot1}
	\end{figure}

An example of such a vine tree structure is shown in Figure \ref{fig:introTreePlot1}.
	It contains the unconditioned bivariate copulas $C_{3,5}$, $C_{1,2}$, $C_{2,3}$ and $C_{3,4}$ in the first tree. 
	Further, it defines the conditional bivariate copulas $C_{2,5;3}$, $C_{1,3;2}$ and $C_{2,4;3}$ in the second tree. Similarly, $C_{4,5;2,3}$, $C_{1,4;2,3}$ are encoded in the third tree and $C_{1,5;2,3,4}$ is defined in the fourth tree.
    
Any regular vine structure, together with the bivariate copulas and marginal distributions, can be uniquely connected to a $d$-dimensional distribution function. 
Let $C_{a_e,b_e;D_e}$ be a bivariate copula associated with the conditional distribution of $(X_{a_e},X_{b_e})$ given $\bm{X}_{D_e}$. Following \cite{BedfordCooke2001} there is a unique $d$-dimensional distribution function $F$, with density given by
\begin{equation}\label{eq:introRVineDensityFormula}
		\begin{split}
		f(x_1,...,x_d) &= \prod\limits_{i=1}^d f_i(x_i)
		 \prod\limits_{i=1}^{d-1} \prod\limits_{e\in E_i} c_{a_e,b_e; D_e}(
		F_{a_e|D_e}(x_{a_e}|\mathbf{x}_{D_e}), F_{b_e|D_e}(x_{b_e}|\mathbf{x}_{D_e})|
		\mathbf{x}_{D_e}).
		\end{split}
	\end{equation}
The associated distribution is called a R-vine distribution. If the R-vine tree structure only consists of paths or star shapes, the subclasses of D-vine or C-vine distributions occur, respectively.
When all marginal densities are the uniform density, then we speak of a regular vine copula. To estimate this copula, \cite{aas2009pair} developed a sequential estimation approach, which remains tractable in high dimensions under a simplifying assumption. In particular we require the following equality:
\begin{equation}\label{eq:introSimplifyingAssumption}
	c_{a_e,b_e;D_e}(w_{a_e},w_{b_e} |\mathbf{x}_{D_e}) = c_{a_e,b_e;D_e}(w_{a_e}, w_{b_e}) \ \forall (w_{a_e}, w_{b_e}) \in [0,1]^2.
\end{equation}
 The simplifying assumption does not assume that the conditional dependence of  $X_{a_e}$ and $X_{b_e}$ given $\mathbf{x}_{D_e}$ is independent of $\mathbf{x}_{D_e}$, since 
\begin{equation}\label{eq:introExampleSimplifyingAssumption}
	F_{a_e,b_e|D_e}(x_{a_e},x_{b_e}|\mathbf{x}_{D_e}) = C_{a_e,b_e;D_e}(
	F_{a_e|D_e}(x_{a_e}|\mathbf{x}_{D_e}),
	F_{b_e|D_e}(x_{b_e}|\mathbf{x}_{D_e}))
\end{equation}
(Theorem 5.15 in \cite{Czado}), meaning $\mathbf{x}_{D_e}$ appears in the arguments $F_{a_e|D_e}(x_{a_e}|\mathbf{x}_{D_e})$ and $F_{b_e|D_e}(x_{b_e}|\mathbf{x}_{D_e})$ of the copula. 
Additionally, as discussed in \cite{nagler2025simplified}, the simplifying assumption does not imply that all conditional dependencies are constant, but only that the dependencies of those copulas specified in the regular vine structure are constant. There has been considerable debate on how restrictive this assumption is and reliable estimation methods for non-simplified vines are only available in small dimension. See \cite{nagler2025simplified} and the reference therein for the current state of affairs. Later, we will apply noise contrastive estimation principles to obtain a non-parametric correction for each observation to simplified vine fits in high dimensions.

\section{Noise contrastive estimation}
\label{sec:NCE}
Noise contrastive estimation (NCE) is a technique used in machine learning to estimate parameters of probabilistic models using supervised learning. \citet[Chapter 14.2.4, pp. 495–497]{hastie2009elements}
already noted that density estimation, which is an unsupervised learning problem, can be performed by
logistic regression, that is, supervised learning. This idea was expanded to problems where computing the normalization constant (or partition function) is expensive or intractable by \citet{Gutmann2012}. 
It is the core of word embeddings such as word2vec \citep{mikolov2013distributed}, generating realistic data based on generative
adversarial networks (GANs) \citep{goodfellow2014generative}, and learning compact and informative representations of high-dimensional data (such as audio, images, or text) without requiring explicit labels using contrastive predictive coding \citep{oord2018representation} to name a few very successful applications. 

\cite{Gutmann2012} start with an unknown multivariate density $g_{true}(\cdot)$  of a random vector 
$\bm Z=(Z_1,\ldots, Z_d)$ with data
${\mathcal{Z}}=\{(z_{i1},\cdots,z_{id}), \linebreak i=1,\ldots,n_{true}\}$. Instead of directly estimating 
$g_{true}$, they reframe the problem as a binary classification task by using the observed data ${\mathcal{Z}}$ and noise data $\mathcal{Z}_{noise}$ containing $n_{noise}$ sampled values from a chosen noise density $g_{noise}$. Here $g_{noise}$ is chosen so that $\mathcal{Z}_{noise}$ can be easily sampled. 
Each observation in $\mathcal{Z}$ is attached with the fictitious label $\gamma=1$ and each observation  in $\mathcal{Z}_{noise}$ with the label $\gamma=0$.
Since $g_{true}$ is unknown, they model $g_{true}$ by a parametric model $g_{model}(\cdot;\boldsymbol{\eta})$ assuming that $g_{true}$ belongs to this family, i.e. there exists a $\boldsymbol{\eta}^*$, which satisfies $g_{true}(\cdot)=g_{model}(\cdot;\boldsymbol{\eta}^*)$. This implies the class conditional probability densities $h(\bm{z}|\gamma=1;\boldsymbol{\eta})=g_{model}(\bm{z};\boldsymbol{\eta})$ and $h(\bm{z}|\gamma=0)=g_{noise}(\bm{z})$.

Next, a binary classification algorithm is used to learn the labels $\gamma\in\{0,1\}$ conditional on $\bm{z}$. For prior probabilities of sampling from each model given by 
\begin{equation}
\label{eq:NCEclassPriorProba1}
P(\gamma=1) = \frac{n_{true}}{n_{true}+n_{noise}} \text{ and }  
P(\gamma=0) = 1-P(\gamma=1),
\end{equation}
Bayes' rule gives posterior probabilities
\begin{equation} \label{eq:NCEclassPosteriorProbabilitiesDef}
	P(\gamma=1|\bm{z};\boldsymbol{\eta}) = \frac{g_{model}(\bm{z};\boldsymbol{\eta})}{g_{model}(\bm{z};\boldsymbol{\eta}) + \nu \cdot g_{noise}(\bm{z})}
    \end{equation} 
and $P(\gamma=0|\bm{z};\boldsymbol{\eta}) = 1-P(\gamma=1|\bm{z}; \boldsymbol{\eta})$ with
$\nu := \frac{n_{noise}}{n_{true}}$. 

Now a probabilistic classifier is used to learn $P(\gamma=1|\bm{z};\boldsymbol{\eta})$ instead of learning $g_{model}(\cdot;\boldsymbol{\eta})$ directly. The corresponding estimator for $P(\gamma=1|\bm{z}; \boldsymbol{\eta})$ is denoted by $k(\bm{z};\boldsymbol{\eta})$. \cite{Gutmann2012} used as estimator $\hat{\boldsymbol{\eta}}$ for 
$\boldsymbol{\eta}$ the value of $\boldsymbol{\eta}$ which maximizes 

\begin{equation}\label{eq:NCEObjectiveFunctionDefinition}
	J(\boldsymbol{\eta}) := \frac{1}{n_{true}} \left( 
	\sum\limits_{\bm{z} \in {\cal{Z}}} \ln(k(\bm{z};\boldsymbol{\eta})) + \sum\limits_{\bm{z} \in {\cal{Z}}_{noise}}  \ln(1-k(\bm{z};\boldsymbol{\eta}))
	\right).
\end{equation}
Assuming that the class labels $\{\gamma_z, z\in {\cal{Z}} \cup {\cal{Z}}_{noise} \}$ are i.i.d Bernoulli distributed and using \eqref{eq:NCEclassPosteriorProbabilitiesDef}, \cite{Gutmann2012} showed that the associated conditional  log-likelihood $l(\boldsymbol{\eta}; \mathbf{z}, \boldsymbol{\gamma})$ 
can be expressed as
\begin{equation}
	\begin{split}
	l(\boldsymbol{\eta}; \bm z, \boldsymbol{\gamma}) & =\sum\limits_{\bm{z} \in {\cal{Z}} \cup {\cal{Z}}_{noise}}\left[\gamma_z \cdot \ln(p(\gamma_z=1|\mathbf{z};\boldsymbol{\eta})) + (1-\gamma_z) \cdot \ln(p(\gamma_z=0|\mathbf{z};\boldsymbol{\eta}))\right] \\
	& = \sum\limits_{\bm{z} \in {\cal{Z}}} \ln(k(\mathbf{z};\boldsymbol{\eta})) + \sum\limits_{\bm{z} \in {\cal{Z}}_{noise}} \ln(1-k(\mathbf{z};\boldsymbol{\eta})) = n_{true} J(\boldsymbol{\eta}),
	\end{split}
\end{equation}
which motivates the objective function $J(\boldsymbol{\eta})$.
Further, they establish a connection between the log-likelihood ratio of model $g_{model}$ to model $g_{noise}$, defined as $G(\bm z; \boldsymbol{\eta})= \ln(g_{model}(\bm z; \boldsymbol{\eta}))-\ln(g_{noise}( \bm z))$, and $k(\cdot;\boldsymbol{\eta})$, which is given by 
\begin{equation}
\label{eq:logratio}
    k(\mathbf{z};\boldsymbol{\eta})= \sigma_{\nu}(G(\bm z; \boldsymbol{\eta})), 
\end{equation}
where $\sigma_{\nu}(x)=[1+\nu \exp (-x)]^{-1}$ is a variant of the sigmoid activation function.
From the definition of $G(\bm z; \boldsymbol{\eta})$ and \eqref{eq:logratio} it follows that $g_{model}$ can be expressed as 
\begin{equation}\label{g_val_eq}
	\begin{split}
	g_{model}(\mathbf{z};\boldsymbol{\eta}) &=\exp(G(\bm z; \boldsymbol{\eta})) \cdot g_{noise}(\bm z) \\
	&= \nu \frac{k(\bm z;\boldsymbol{\eta})}{ 1-k(\bm z;\boldsymbol{\eta})} \cdot g_{noise}(\bm z).
	\end{split}
\end{equation}
Since $g_{noise}$ is known and $\boldsymbol{\eta}$ is estimated by $\hat{\boldsymbol{\eta}}$, an estimate for  $g_{model}$ is readily obtained.

 Under regularity conditions (a)-(c) given in Theorem 2 of \cite{Gutmann2012}, consistency and asymptotic normality of the estimator $\hat{\boldsymbol{\eta}}$ were shown. It is worth noting that in their approach, no normalization constraint for $g_{model}(\cdot;\boldsymbol{\eta})$ is required.
Early approaches used a logistic classifier, but modern machine-learning-based classifiers can also be employed, as proposed here. 
  
\section{Improving the fit of simplified vine copulas by using noise contrastive estimation}
\label{sec:vine-NCE}
As already mentioned, we want to apply the noise contrastive estimation approach to make an observation-specific adjustment to a simplified vine copula fit. For this, we start with pseudo copula data as defined in \eqref{eq:pseudo}. We collect them in the set ${\mathcal{U}}=\{(u_{i1},\cdots,u_{id}),i=1,\ldots,n_{true}\}$.
The copula data $\mathcal{U}$ follows an unknown $d$-dimensional copula distribution $C_{true}$ with copula density $c_{true}$ we want to estimate, this means $c_{true}$ corresponds to $g_{true}$ in Section \ref{sec:NCE}. As parametric model for $c_{true}$ we choose a neural-network-based density $c_{model}(\cdot;\boldsymbol{\eta})$ for $c_{true}$ with parameter vector $\boldsymbol{\eta}$ corresponding to the layer weights of the neural network used to for the probabilistic classifier. Again, we make the assumption that $c_{true}$ can be estimated by $c_{model}(\cdot; \boldsymbol{\eta}^{*})$, i.e. that there exists a parameter vector $\boldsymbol{\eta}^{*}$, such that $c_{model}(\cdot; \boldsymbol{\eta}^{*}) = c_{true}(\cdot)$. 
For the choice of the noise distribution we fit a simplified regular vine copula model to $\mathcal{U}$ using the algorithm by \cite{Dissmann2013} implemented in the R-package \texttt{rvinecopulib} by \cite{rvinecopulib} and use this fitted copula distribution. We denote the fitted simplified copula density by $c_{noise}$ and simulate $n_{noise}$ i.i.d. vectors from $c_{noise}$ to form a dataset ${\mathcal{U}_{noise}}$.

We proceed as in Section \ref{sec:NCE} using the sample $\mathcal{U} \cup \mathcal{U}_{noise}$ to train a neural-network-based probabilistic classifier to estimate 
$k(\cdot; \boldsymbol{\eta})$. A slight modification to the objective function \eqref{eq:NCEObjectiveFunctionDefinition} is made by changing the denominator $n_{true}$ to $n_{true}+n_{noise}$. This allows to interpret the objective function as negative binary cross entropy loss known in machine learning.
In Section \ref{sec:sim}, through a simulation study, we assess the ability of this trained classifier in identifying the correct class labels and estimating the unknown copula density $c$.

Having estimated $k(\bm u;\boldsymbol{\eta})$ and using the known simplified vine copula density $c_{noise}$, we can estimate $c_{model}$ similarly to \eqref{g_val_eq}.
Next, we argue that the assumption
$c_{model}(\cdot; \boldsymbol{\eta}^{*}) = c_{true}(\cdot)$ is reasonable in this context.
Note that $k(\cdot; \boldsymbol{\eta})$ is a neural network with sigmoid activation function in the output layer. In the universal approximation theorem for neural networks of \cite{Hornik1991}, no activation function is assumed.
Therefore, letting $\tilde{k}(\cdot;\boldsymbol{\eta})$ denote the output of the neural network on the logit level, i.e. the outputs of a neural network without activation function:
\begin{equation}\label{eq:ReasoningCorrectModelLogitLevel}
	k(\mathbf{u}; \boldsymbol{\eta}) = \frac{1}{1+\exp(-\tilde{k}(\mathbf{u};\boldsymbol{\eta}))} \Leftrightarrow \tilde{k}(\mathbf{u};\boldsymbol{\eta}) = \ln \left( \frac{k(\mathbf{u}; \boldsymbol{\eta})}{1-k(\mathbf{u}; \boldsymbol{\eta})} \right) \forall \mathbf{u} \in [0,1]^d.
\end{equation}
Then for all $\mathbf{u} \in [0,1]^d$ it holds that 
\begin{equation}\label{eq:ReasoningCorrectModelCTrueCModelEquality}
	\begin{split}
	c_{true}(\mathbf{u}) = c_{model}(\mathbf{u};\boldsymbol{\eta}^*) &\overset{\text{(\ref{g_val_eq})}}{\Leftrightarrow} c_{true}(\mathbf{u}) =\nu \frac{k(\mathbf{u};\boldsymbol{\eta}^*)}{1-k(\mathbf{u};\boldsymbol{\eta}^*)} \cdot c_{noise}(\mathbf{u}) \\
&\overset{\text{(\ref{eq:ReasoningCorrectModelLogitLevel})}}{\Leftrightarrow}
	c_{true}(\mathbf{u}) = \nu \exp(\tilde{k}(\mathbf{u};\boldsymbol{\eta}^*)) c_{noise}(\mathbf{u}) \\
	&\overset{\phantom{\text{(\ref{g_val_eq})}}}{\Leftrightarrow}
	\tilde{k}(\mathbf{u};\boldsymbol{\eta}^*) = \ln \left( \frac{c_{true}(\mathbf{u})}{\nu c_{noise}(\mathbf{u})} \right) 
	\end{split}
\end{equation}
Assuming that $c_{true}(\cdot)$ is continuous, and using that $c_{noise}(\cdot)$ is continuous, also $\ln \left( \frac{c_{true}(\cdot)}{\nu c_{noise}(\cdot)} \right)$ is continuous. Further, $\tilde{k}(\cdot;\boldsymbol{\eta})$ is the output of a neural network without activation function. Hence, the universal approximation theorem is applicable and gives that for any $\epsilon > 0$, $\ln \left( \frac{c_{true}(\cdot)}{\nu c_{noise}(\cdot)} \right) \colon [0,1]^d \to \mathbb{R}$ can be approximated arbitrarily well by a neural network with sufficiently many hidden units on the compact interval $[\epsilon, 1-\epsilon]^d$. Thus, if $k(\cdot;\boldsymbol{\eta})$, and thus also $\tilde{k}(\cdot; \boldsymbol{\eta})$, have sufficiently many hidden units, it can be concluded, that $c_{true}(\cdot)$ can be approximated arbitrarily well by $c_{model}(\cdot;\boldsymbol{\eta})$ for some parameter vector $\boldsymbol{\eta}$, on the compact interval $[\epsilon, 1-\epsilon]^d$, by Equation (\ref{eq:ReasoningCorrectModelCTrueCModelEquality}).
\section{Illustration and simulation study}
\label{sec:sim}

\subsection{Illustration of the correction}

We illustrate the approach for $d=5$ by using non-simplified vine copula data based on the vine tree structure of Figure \ref{fig:introTreePlot1}. Section 3.1 of \cite{kraus25} shows how to simulate from a non-simplified vine copula, where the copula parameters of each pair copula in Tree 2, ..., Tree $d-1$ are changing. The specific choices of the pair copula families and the functions which change the copula parameter are specified in the appendix and visualized in the left panel of Figure \ref{fig:chap2sim1} for $n_{true}=10000$. The non-simplified nature of the data is visible by the non-elliptical shapes of the contours. A simplified vine copula restricted to only parametric pair copulas is fitted to this data. The fitted parametric families and their estimated copula parameters are given in Table \ref{tab:fittedSimplifiedVineExample} in the appendix. Now $n_{noise}=10000$ samples from this simplified vine copula model are drawn and visualized in the right panel of Figure \ref{fig:chap2sim1}. Clearly, the simplified model is not appropriate for this data, since both panels of Figure \ref{fig:chap2sim1} do not show the same patterns.

\begin{figure}[t]
	\centering
	\includegraphics[width=0.4\textwidth ]{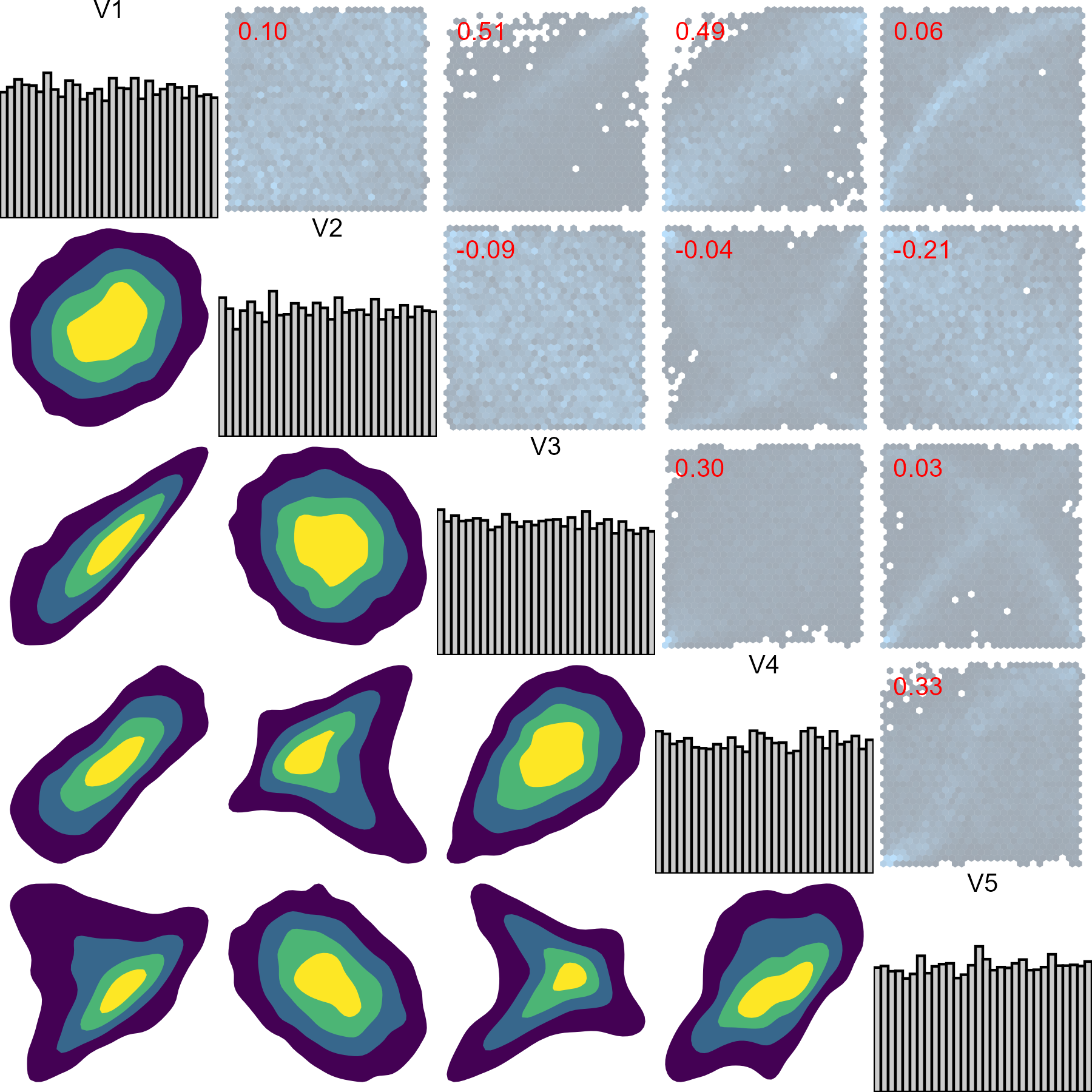}
    \hspace*{.5cm}
    \includegraphics[width=0.4\textwidth ]{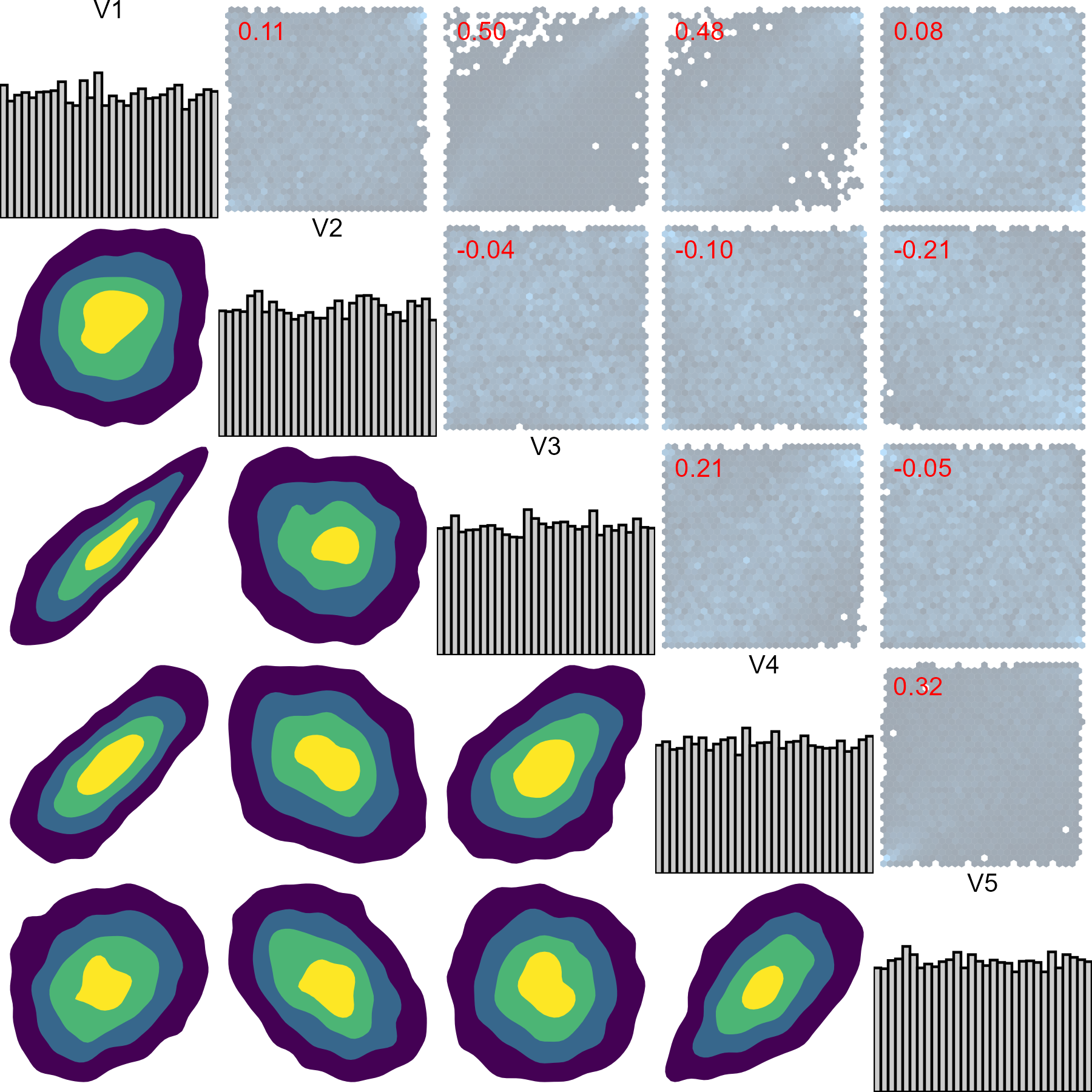}
	\caption{Left panel: Five dimensional non-simplified vine copula data (upper triangular), along with marginally $N(0,1)$ contour plots (lower triangular) and marginal histograms (diagonal)
    Right panel: 10000 samples from fitted simplified vine copula model based on the non-simplified data of the right panel.}
	\label{fig:chap2sim1}
\end{figure}

Next, we construct a neural–network–based probabilistic classifier for the
combined sample consisting of data generated from the non-simplified model and
from the fitted simplified model. The neural network is implemented in \textit{R}
using the packages \textit{keras} (\cite{RPackageKeras}) and \textit{tensorflow}
(\cite{RPackageTensorflow}).
The architecture comprises two hidden layers with $32$ and $16$ units,
respectively. Both hidden layers use the leaky ReLU activation function, $
\operatorname{leaky\_relu}(z) = \max(z, \alpha z), \  \alpha = 0.1,
$ the output layer consists of a single unit with a sigmoid activation
function. The network is trained using the \textit{adam}
optimizer (\cite{kingma2014}). Training is performed for $200$ epochs with a batch
size of $128$. The learning rate is initialized at $0.01$ and is reduced by
half every $30$ epochs.
The data set is split as follows: $20\%$ of the full data set is held out as a test
set and not used during training. Of the remaining $80\%$, an additional $20\%$
is used as validation data during training, while the remainder constitutes the
training set. The evolution of training and validation accuracy over the epochs
is shown in Figure~\ref{fig:chap4NN1} in Appendix~\ref{sec: Appendix A}.

The final training accuracy reaches approximately $87\%$, with the validation
accuracy stabilizing slightly lower at around $85\%$. Evaluation on the
independent test set yields an accuracy of $84.58\%$. Hence, for this data set,
the classifier correctly distinguishes about $85\%$ of the samples, closely
matching the performance observed on the validation set. The classifier does not
exhibit signs of overfitting and successfully learns to differentiate between
the original and the simulated simplified samples.

\begin{figure}[t]
	\centering
	\includegraphics[width=.8\textwidth ]{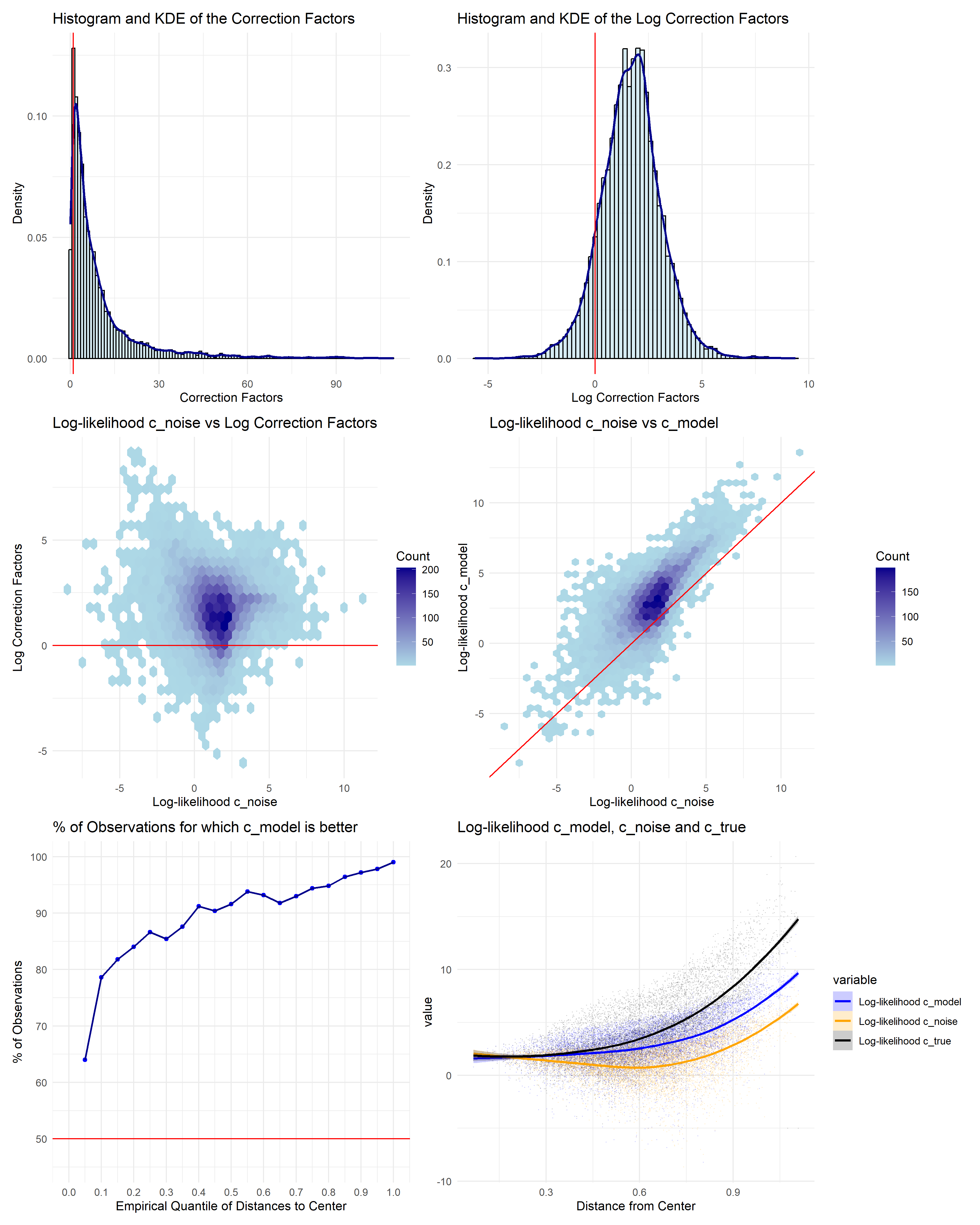}
	\caption{Top row shows histogram and kernel density estimate of the correction factors $\frac{k(\mathbf{u}_i;\boldsymbol{\hat{\eta}})}{1-k(\mathbf{u}_i;\boldsymbol{\hat{\eta}})}$ and the log correction factors. Second row shows log-likelihoods $\ln(c_{noise}(\mathbf{u}_i))$ on the x-axis against the log correction factors on the left and log-likelihoods $\ln(c_{model}(\mathbf{u}_i;\hat{\boldsymbol{\eta}}))$ on the right. Bottom row shows the empirical quantiles of the distance from the center, $||\mathbf{u}_i - 0.5 \cdot \mathbf{1}_5||_{2}^2$, where $\mathbf{1}_5 = (1,...,1)^T \in \mathbb{R}^5$ against the percentage of observations for which $c_{model}(\mathbf{u}_i;\hat{\boldsymbol{\eta}}) > c_{noise}(\mathbf{u}_i)$ on the left, and the distance from the center against the log-likelihoods of both models and the true underlying distribution $c_{true}$ on the right.}
	\label{fig:AllPlotsInOneLinearExample}
\end{figure}

A histogram and kernel density estimate of the correction factors $\frac{k(\mathbf{u}_i;\boldsymbol{\eta})}{1-k(\mathbf{u}_i;\boldsymbol{\eta})}$  for $\nu=1$ of \eqref{g_val_eq} is shown in the top left of Figure \ref{fig:AllPlotsInOneLinearExample}, the highest $2\%$ of observations are removed for better visibility. The top right plot shows the log correction factors, $\ln \left( \frac{k(\mathbf{u}_i;\boldsymbol{\eta})}{1-k(\mathbf{u}_i;\boldsymbol{\eta})} \right)$, where none of the observations have been removed. 
The second row in Figure \ref{fig:AllPlotsInOneLinearExample} shows a plot of the log-likelihood under the simplified vine copula, $\ln(c_{noise}(\mathbf{u}_i))$, versus the log correction factors $\ln \left( \frac{k(\mathbf{u}_i;\boldsymbol{\hat{\eta}})}{1-k(\mathbf{u}_i;\boldsymbol{\hat{\eta}})} \right)$ on the left, while the second row on the right contains a plot of the log-likelihoods under the simplified vine copula $\ln(c_{noise}(\mathbf{u}_i))$ against the log-likelihoods under the model derived via noise contrastive estimation, $\ln(c_{model}(\mathbf{u}_i; \boldsymbol{\hat{\eta}}))$.  
In the bottom right of Figure \ref{fig:AllPlotsInOneLinearExample} the distance of the observations $\mathbf{u}_i$ from the center of the unit cube, $||\mathbf{u}_i - 0.5 \cdot \mathbf{1}_5||_2$, with $\mathbf{1}_5 = (1,...,1)^T \in \mathbb{R}^5$ is plotted on the x-axis against the log-likelihoods $\ln(c_{noise}(\mathbf{u}_i))$, $\ln(c_{model}(\mathbf{u}_i; \hat{\boldsymbol{\eta}}))$ and $\ln(c_{true}(\mathbf{u}_i))$. Points are individual observations, while lines are smoothed. This indicates, that the model density is higher mainly in the outer parts of the unit hypercube, so most of the correction takes place in the tails, not in the center of the distribution. Especially there, $c_{model}(\cdot;\hat{\boldsymbol{\eta}})$ is much closer than $c_{noise}$ to the true distribution $c_{true}$. Similarly, the plot in the bottom left groups the data points $\mathbf{u}_i$ into 20 groups of the same size, corresponding to the empirical $5\%$ quantile levels of the distances to the center $||\mathbf{u}_i - 0.5 \cdot \mathbf{1}_5||_2$. For instance, the group corresponding to the $5\%$ quantile contains all observations $\mathbf{u}_i \in [0,1]^5$, that are among the $\frac{10000}{20} = 500$ observation vectors with the lowest values of $||\mathbf{u}_i - 0.5 \cdot \mathbf{1}_5||_2$. On the y-axis the percentage of observations $\mathbf{u}_i$ in the respective groups are shown, for which $c_{model}(\mathbf{u}_i; \hat{\boldsymbol{\eta}}) > c_{noise}(\mathbf{u}_i)$, i.e. for which the model derived via noise contrastive estimation assigns higher likelihoods than the simplified vine copula model. This plot again shows that $c_{model}(\cdot; \hat{\boldsymbol{\eta}})$ captures the structure of the data better than $c_{noise}$ especially further away from the center, i.e. in the tails. For the highest quantiles of $||\mathbf{u}_i - 0.5 \cdot \mathbf{1}_5||_2$ almost every point is assigned a higher likelihood under the noise contrastive estimation model than under the simplified vine copula model. For the lower quantiles this is only true for about $65 \%$ to $80 \%$ of the samples. 

\clearpage 
\subsection{Simulation study}
We simulate using the vine tree structure of Figure \ref{fig:introTreePlot1} and allowing the pair copula parameters of Tree 2 to Tree 4 to depend on the conditioning values in a linear, quadratic and cubic fashion. We investigate three different dependence scenarios for the range of Kendall's $\tau$ (weak: $\tau \in (.001, .3)$, medium: $\tau \in (.001,.6)$, strong: $\tau \in (.001,.9)$). The estimated Jensen-Shannon divergence (JSD, \cite{lin2002divergence}) between the true model $c_{true}$ and the noise model $c_{noise}$ for each scenario shows that the JSD increases with the dependence level and is largest for the quadratic and cubic variation of the parameters (Table \ref{tab:simStudyDistributionDistances} in Appendix \ref{sec: Appendix B}). Further details on the underlying true model $c_{true}$ can be found in Appendix \ref{sec: Appendix B}.

For fitting the noise model $c_{noise}$, i.e., the simplified vine, we allowed all 
pair-copula families implemented in \texttt{rvinecopulib}, including both parametric 
and non-parametric options. For $n_{true}=1000$ we generated $n_{noise}=10000$ samples 
from the fitted simplified vine, and for $n_{true}=10000$ we used $n_{noise}=40000$.
The neural network used in $c_{model}(\cdot; \boldsymbol{\hat{\eta}})$ has two hidden layers with 20 and 10 units, 
respectively, using leaky-ReLU activation (slope 0.1 on $(-\infty,0)$). The output 
layer consists of a single sigmoid unit. The model is trained with the Adam optimizer 
for 200 epochs, batch size 128, and an initial learning rate of 0.01, which is halved 
every 30 epochs.

\begin{figure}[t]
	\centering
	\includegraphics[width=\textwidth ]{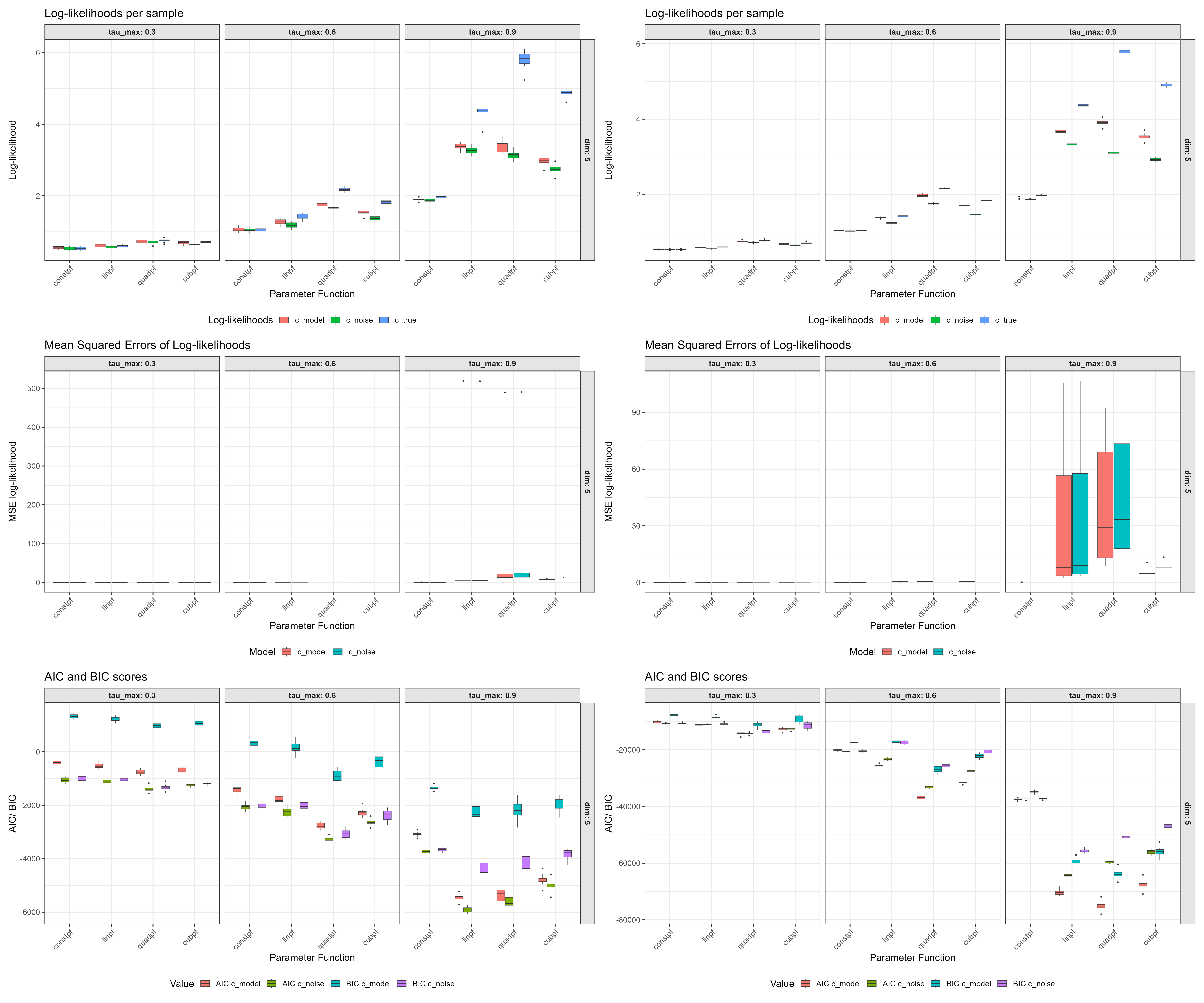}
	\caption{
    Boxplots of log-likelihoods per sample (first row) , mean square errors with regard to $c_{true}$ (second row),  AIC/BIC scores (third row) based 1000 samples (left column) and 10000 samples (right column) over 10 replications. 
    }
	\label{fig:simulation_final_cropped}
\end{figure}

The results for the log-likelihood per sample, the mean squared error (MSE) of simplified vine copula model $c_{noise}$ with 
respect to $c_{true}$,
\[
\frac{1}{n_{true}} \sum_{i=1}^{n_{true}}
\left( \ln c_{true}(\mathbf{u}_i)
     - \ln c_{noise}(\mathbf{u}_i) \right)^2,
\]
and analogously for the corrected model $c_{model}(\cdot;\hat{\boldsymbol{\eta}})$, 
as well as the AIC/BIC scores, are shown in 
Figure~\ref{fig:simulation_final_cropped}.
The difference in log-likelihood per sample between $c_{model}(\cdot ; \boldsymbol{\hat{\eta}})$ and $c_{true}$ 
increases with dependence strength and is largest under quadratic variation, 
followed by cubic and linear variation.  
The same pattern holds for $c_{noise}$, although $c_{model}(\cdot ; \boldsymbol{\hat{\eta}})$ remains consistently 
closer to $c_{true}$.  
Larger sample sizes reduce both the deviation from $c_{true}$ and the variability, 
as expected.  
Overall, $c_{model}(\cdot ; \boldsymbol{\hat{\eta}})$ provides a clearly better fit than the simplified vine model 
$c_{noise}$.
For the MSE relative to $c_{true}$, the scenario with quadratic variation and strong 
dependence produces the highest errors for both sample sizes, while the remaining 
scenarios exhibit similar MSE values.  
Again, $c_{model}(\cdot ; \boldsymbol{\hat{\eta}})$ yields lower MSE than $c_{noise}$.
Regarding AIC/BIC, $c_{model}(\cdot ; \boldsymbol{\hat{\eta}})$ attains larger values due to the additional parameters 
introduced by the neural network correction.  
The scores decrease (indicating improved fit) as the dependence range increases. The average AIC/BIC scores are also given in Table \ref{tab:all_noise_simulation_study_AIC_BIC} in Appendix \ref{sec: Appendix B}.
Since AIC/BIC are not normalized per observation, they are substantially lower for 
the larger sample size. Overall, we see that the NCE based non-parametric correction  $c_{model}(\cdot ; \boldsymbol{\hat{\eta}})$ improves the  model fit over the simplified vine copula model $c_{noise}$ especially when the level of the varying dependence strength is large. Additional simulations for $d=3$ can be found in \cite{kraus25} in Chapter 5, giving the same conclusions.

\clearpage

\section{Applications}

\label{sec:app}


\subsection{Abalone Data}\label{sec:ApplicationAbalone}

The Abalone data (\cite{AbaloneData}) has $4177$ observations of nine
variables. We use seven of them: \textit{Length} (longest shell measurement), \textit{Diameter}
(perpendicular shell measure), \textit{Height}, \textit{Whole\_weight}, \textit{Shucked\_weight},
\textit{Viscera\_weight}, and \textit{Shell\_weight} (dry weight). Before fitting 
models, the data is transformed to the copula scale using kernel density estimation as implemented in \textit{kde1d} (\cite{RPackageKDE1D}). Since \textit{Height} has only 51
distinct values, we apply jittering following \cite{NaglerKernelPaper1,
NaglerKernelPaper2} 

The left panel of Figure~\ref{fig:PairPlotOrigAbaloneExample} (Appendix~\ref{sec: Appendix C})
shows pairwise scatter plots and marginally normalized contour plots of the pseudo-copula data.
A simplified vine copula with parametric and non-parametric families is fitted to the data,
with results given in Table~\ref{tab:fittedSimplifiedVineAbaloneExample}. 
From this model, $5 \cdot 4177 = 20885$ samples are simulated using 
$\nu = n_{noise}/n_{true} = 5$ to provide sufficient training data. 
The simulated data, shown alongside the original in Figure~\ref{fig:PairPlotOrigAbaloneExample},
appears visually similar, suggesting that the simplified vine copula fits the data well.

A neural network with 20 and 10 units in its two hidden layers is trained using the same
settings as in the $d=5$ example to distinguish original data from simplified-vine samples.
The data are split into $80\%$ training (with $20\%$ of that used for validation) and $20\%$ testing.
After training, the binary cross-entropy loss is 0.4378 (accuracy $83.36\%$) on the training set
and 0.4379 (accuracy $83.32\%$) on the test set, indicating no overfitting.



The neural network contains 381 parameters, while the simplified vine copula has
466.1615 parameters. The latter is non-integer because, for
non-parametric copulas, the effective number of parameters is computed via the
trace-based estimator of \cite{SchellhaseKauermann2014}.

The model $c_{noise}$ yields an average copula log-likelihood per sample of $10.186$, an AIC of $-84159.418$, and a BIC of $-81205.19$. In contrast, $c_{model}(\cdot;\hat{\boldsymbol{\eta}})$ attains a slightly higher copula log-likelihood of $10.268$ per sample and produces a worse information criteria, with AIC $-84087.234$ and BIC $-78718.476$, i.e. $c_{noise}$ has a better fit with regard to AIC/BIC than $c_{model}(\cdot ; \boldsymbol{\hat{\eta}})$, thus the simplified vine model is preferred over the corrected one as expected.

\begin{figure}[h]
	\centering
	\includegraphics[width=.6\textwidth ]{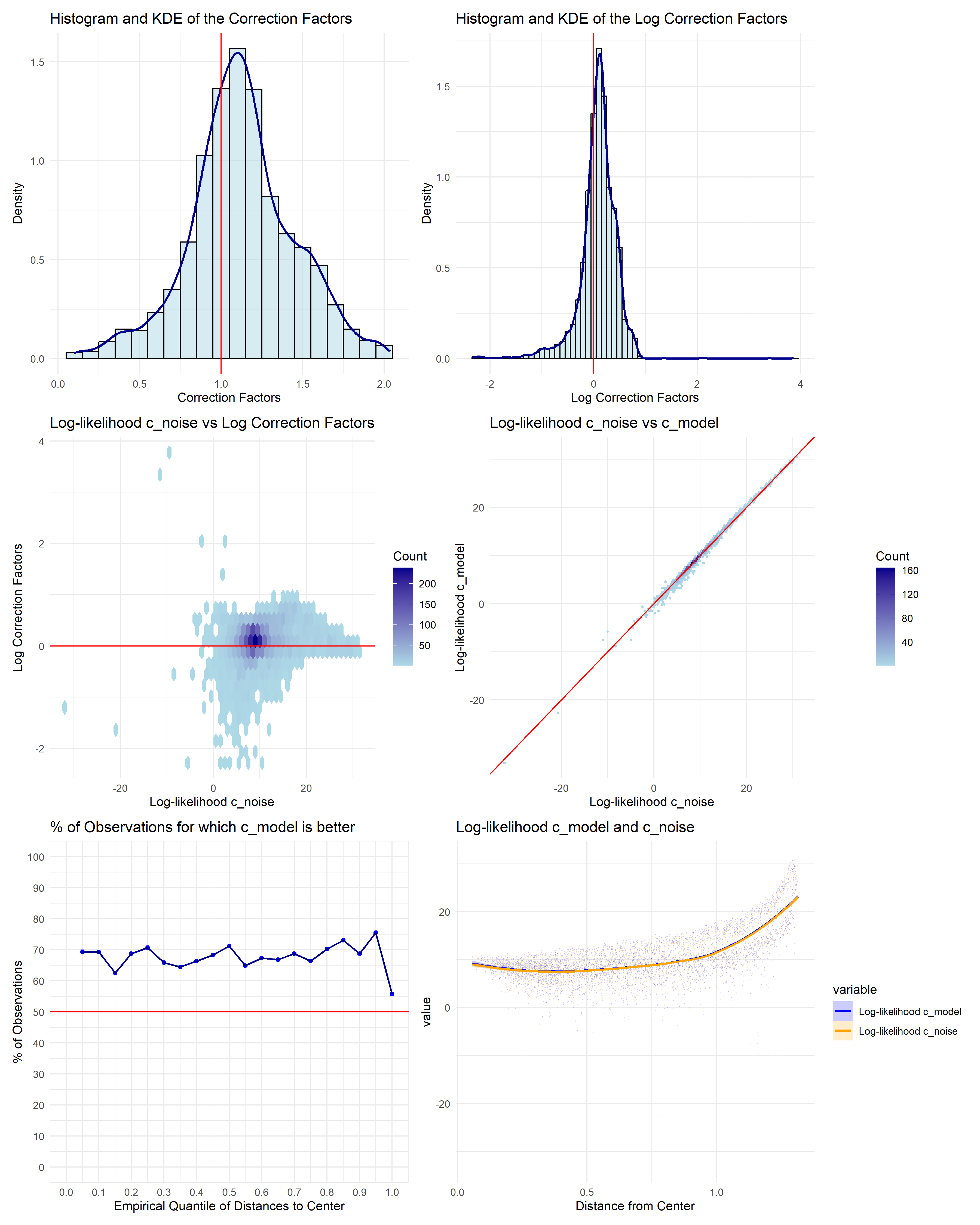}
	\caption{Abalone data:
    The top row shows a histogram and kernel density estimate of the correction factors 
(top $2\%$ removed) and the corresponding log correction factors (no trimming). 
The second row plots $\ln(c_{noise}(\mathbf{u}_i))$ against the log correction factors (left) 
and against $\ln(c_{model}(\mathbf{u}_i;\hat{\boldsymbol{\eta}}))$ (right). 
The bottom row displays empirical quantiles of the distance to the center 
$||\mathbf{u}_i - 0.5\mathbf{1}_7||_2^2$ versus the percentage of points with 
$c_{model}(\mathbf{u}_i;\hat{\boldsymbol{\eta}}) > c_{noise}(\mathbf{u}_i)$ (left), 
and the distance versus the log-likelihoods of both models (right).
    }
	\label{fig:AllPlotsInOneAbaloneExample}
\end{figure}

The top-left panel of Figure~\ref{fig:AllPlotsInOneAbaloneExample} shows the histogram and kernel density estimate of the correction factors (top $2\%$ removed). Most factors lie very close to $1$, indicating that $c_{noise}$ already fits the data well and requires only minor adjustment by the NCE model.
The second-row plots confirm that $c_{model}(\cdot;\hat{\boldsymbol{\eta}})$ and $c_{noise}$ are overall very similar.
To study whether observations in the tails or the center especially benefit from the NCE correction consider the bottom-left panel. 
It groups observations into 20 equally sized bins based on their distance from the center $||\mathbf{u}_i-0.5\mathbf{1}_7||_2$. For each group, it displays the proportion of points for which $c_{model}(\mathbf{u};\hat{\boldsymbol{\eta}}) > c_{noise}(\mathbf{u})$. While the NCE corrected model is preferred in most cases, this plot does not reflect the magnitude of the improvement.
In the bottom-right panel, log-likelihoods $\ln(c_{noise}(\mathbf{u}_i))$ and $\ln(c_{model}(\mathbf{u}_i;\hat{\boldsymbol{\eta}}))$ are plotted against $||\mathbf{u}_i-0.5\mathbf{1}_7||_2$. The smoothed curves nearly coincide, with the one for $c_{model}(\cdot ; \boldsymbol{\hat{\eta}})$ only slightly higher, again showing that both densities behave very similarly.
\clearpage

\subsection{Magic Gamma Telescope Data}\label{sec:ApplicationMagic}

The Magic Gamma Telescope data set (\cite{MagicGammaTelescopeData}) contains 19{,}020
observations with 10 continuous features. After transforming the data to the copula
scale (as in the first application), the resulting pseudo-copula data are shown in
Figure~\ref{fig:PairPlotOrigMagicExample} (Appendix~\ref{sec: Appendix C}). Again, both parametric and
non-parametric bivariate families are allowed when fitting $c_{noise}$. The fitted model is reported in
Table~\ref{tab:fittedSimplifiedVineMagicExample}, and simulated samples are
visualized in Figure~\ref{fig:PairPlotOrigMagicExample} (Appendix~\ref{sec: Appendix C}).

Using $\nu = n_{noise}/n_{true} = 2$, we generate 38{,}040 noise samples, yielding
57{,}060 total observations for training the classifier. The same neural network
architecture as in the Abalone example is used, with a split of $20\%$ test data
and $20\%$ of the remaining data for validation. After training, the binary
cross-entropy loss is 0.5497 (accuracy $70.21\%$) on the training set and
0.5691 (accuracy $69.13\%$) on the test set.

The neural network has 451 
parameters, while $c_{noise}$ has 3782.168 effective parameters.
For $c_{noise}$, the average copula log-likelihood per sample is $6.743$, with AIC $-248927.816$ and BIC $-219225.519$. In comparison, $c_{model}(\cdot;\hat{\boldsymbol{\eta}})$ attains a higher log-likelihood per sample of $7.144$ and yields better information criteria, with AIC $-263306.529$ and BIC $-230140.951$.
This indicates that the NCE corrected model $c_{model}(\cdot;\hat{\boldsymbol{\eta}})$ is a better fit than the simplified vine copula model $c_{noise}$.
So the NCE correction is useful in this application. 
\begin{figure}[h]
	\centering
	\includegraphics[width=.7\textwidth ]{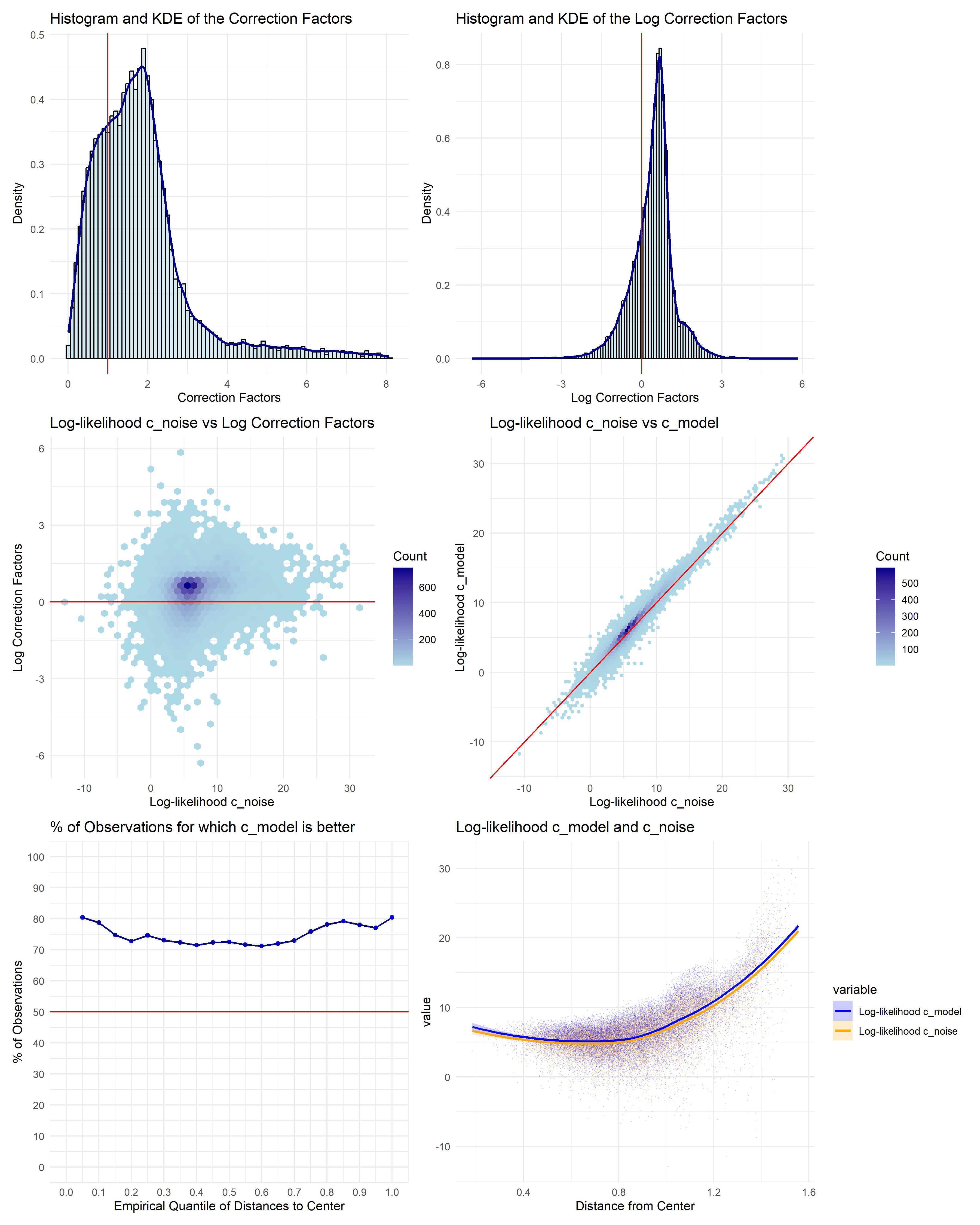}
	\caption{Magic Gamma Telescope data:
    The top row shows a histogram and kernel density estimate of the correction factors 
(top $2\%$ removed) and the corresponding log correction factors (no trimming). 
The second row plots $\ln(c_{noise}(\mathbf{u}_i))$ against the log correction factors (left) 
and against $\ln(c_{model}(\mathbf{u}_i;\hat{\boldsymbol{\eta}}))$ (right). 
The bottom row displays empirical quantiles of the distance to the center 
$||\mathbf{u}_i - 0.5\mathbf{1}_{10}||_2^2$ versus the percentage of points with 
$c_{model}(\mathbf{u}_i;\hat{\boldsymbol{\eta}}) > c_{noise}(\mathbf{u}_i)$ (left), 
and the distance versus the log-likelihoods of both models (right).
    }
	\label{fig:AllPlotsInOneMagicExample}
\end{figure}

The top-left panel of Figure~\ref{fig:AllPlotsInOneMagicExample} shows the histogram and
kernel density of the correction factors (top $2\%$ removed). Most values exceed~1, indicating
that $c_{model}(\cdot;\hat{\boldsymbol{\eta}})$ generally fits the data better than $c_{noise}$.
The second-row plots confirm that $c_{model}(\cdot ; \boldsymbol{\hat{\eta}})$ assigns higher likelihoods overall.
The bottom-left panel groups points into 20 distance-based quantile bins
($||\mathbf{u}_i - 0.5\mathbf{1}_{10}||_2$) and shows that $c_{model}(\cdot ; \boldsymbol{\hat{\eta}})$ yields larger likelihoods
for about $70$--$80\%$ of observations in every group. The bottom-right panel plots the distance
to the center against $\ln(c_{noise}(\mathbf{u}_i))$ and $\ln(c_{model}(\mathbf{u}_i;
\hat{\boldsymbol{\eta}}))$; the smoothed curve for $c_{model}(\cdot ; \boldsymbol{\hat{\eta}})$ lies consistently above that for
$c_{noise}$, again indicating a better fit.

\clearpage

\section{Conclusions and outlook}
\label{sec:summary}

Our calibration method for simplified vine copulas depends on NCE’s ability to distinguish a vine copula’s noise samples from actual observations. As improved discrimination techniques for copula data emerge—such as \cite{huk2026diffusion}—our approach can readily integrate and benefit from these advances.
Another area to utilize this approach is to construct hypothesis tests whether the calibration is needed. For this, an asymptotic likelihood ratio approach might be a starting approach, however since we used a two step estimation approach to estimate the marginal distributions first and then the copula approach, the asymptotic distribution is not straightforward. Thus it might be preferable to utilize a bootstrap approach, which will be a further 
research topic.

\backmatter

\bmhead{Supplementary Information}
The appendix is included in this submission, and the code used for experiments is available at the following repository \url{https://github.com/Michael-K5/Calibrating-Simplified-Vine-Copulas-with-NCE}.

\bmhead{Acknowledgements}
Microsoft Copilot assisted with grammar, clarity, and shortening suggestions, and all changes were verified by the authors.

\bmhead{Author Contributions}
Conceptualization, C.C. and D.H.; Methodology, C.C. and M.D.K.; Writing, C.C. and M.D.K.; Computational Implementation M.D.K.; Supervision, C.C. and D.H.; All authors reviewed the manuscript.

\bmhead{Funding}
Claudia Czado is supported by the German Research Foundation and the TUM International Graduate School of Science and Engineering (IGSSE). David Huk is funded by the Center for Doctoral Training in Mathematical Sciences at the University of Warwick.

\bmhead{Data Availability}
The data and code used for the experiments are available at the following repository: \url{https://github.com/Michael-K5/Calibrating-Simplified-Vine-Copulas-with-NCE}.

\section*{Declarations}
\bmhead{Competing Interests}
The authors declare no competing interests.

\begin{appendices}
\section{Details on 5-dimensional illustration}
\label{sec: Appendix A}

\setcounter{figure}{6}
\renewcommand{\thefigure}{\arabic{figure}}

\setcounter{table}{0}
\renewcommand{\thetable}{\arabic{table}}

\subsection{Specification of non-simplified vine used in the 5-dimensional illustration:}
For the vine structure of  Figure \ref{fig:introTreePlot1}, the pair copulas are specified as 
\begin{itemize}
	\item Tree 1: $\begin{cases}
		\mathbb{C}_{2,5}:\text{ Frank , parameter } -1.861 \ ( \tau=-0.2) . \\
		\mathbb{C}_{3,4}: \text{ Clayton , parameter } 0.857 \ (\tau=0.3) . \\
		\mathbb{C}_{2,3}: \text{ Gaussian , parameter } -0.156 \ (\tau=-0.1) . \\
		\mathbb{C}_{1,2}: \text{ Frank , parameter } 0.907 \ (\tau= 0.1) .
	\end{cases}$
	\item  Tree 2: $\begin{cases} 
		\mathbb{C}_{3,5;2}: \text{ Frank , linpf} ((1), \text{Frank})(u_2) \text{ (see Eq. (\ref{def:linearParameterFunction})) }. \\
		\mathbb{C}_{2,4;3}: \text{ Gaussian , linpf} ((1), \text{Gauss})(u_3) . \\
		\mathbb{C}_{1,3;2}: \text{ Joe , linpf} ((1), \text{Joe})(u_2) . \end{cases}$
	\item Tree 3: $\begin{cases}
		\mathbb{C}_{1,5;2,3}: \text{ Gaussian , linpf} ((0.7,0.3)^T, \text{Gauss})(u_2,u_3) . \\
		\mathbb{C}_{1,4;2,3}: \text{ Gumbel , linpf} ((0.4,0.6)^T, \text{Gumbel})(u_3,u_2) . \end{cases}$
	\item Tree 4: $\mathbb{C}_{4,5;1,2,3}$: Gaussian , linpf$((0.2,0.5,0.3)^T, \text{Gauss})(u_2,u_3,u_1)$.
\end{itemize}
where
\begin{equation}\label{def:linearParameterFunction}
	\begin{split}\text{linpf}(\mathbf{a}, \text{family})(\mathbf{u}) &= \textup{ktau\_to\_par(family, tanh($(T_{max} - T_{min}) \cdot arg + T_{min}$))}, \\
\text{where }arg &=\mathbf{a}^T \cdot \mathbf{u} = \sum_i a_i \cdot u_i \\
\text{and } T_{min} &=
\begin{cases} \textup{artanh}(0.001) & \text{if family is Clayton, Gumbel or Joe } \\
	\textup{artanh}(\tau_{min}) & \text{else, where } \tau_{min} = -0.92
\end{cases}\\
\text{and } T_{max} &= \textup{artanh}(\tau_{max}), \text{ with } \tau_{max} = 0.92.
\end{split}
\end{equation}
\clearpage
\subsection{Performance over epochs in the 5-dimensional neural network illustration:}
\begin{figure}[h]
	\centering
	\includegraphics[width=0.7\textwidth ]{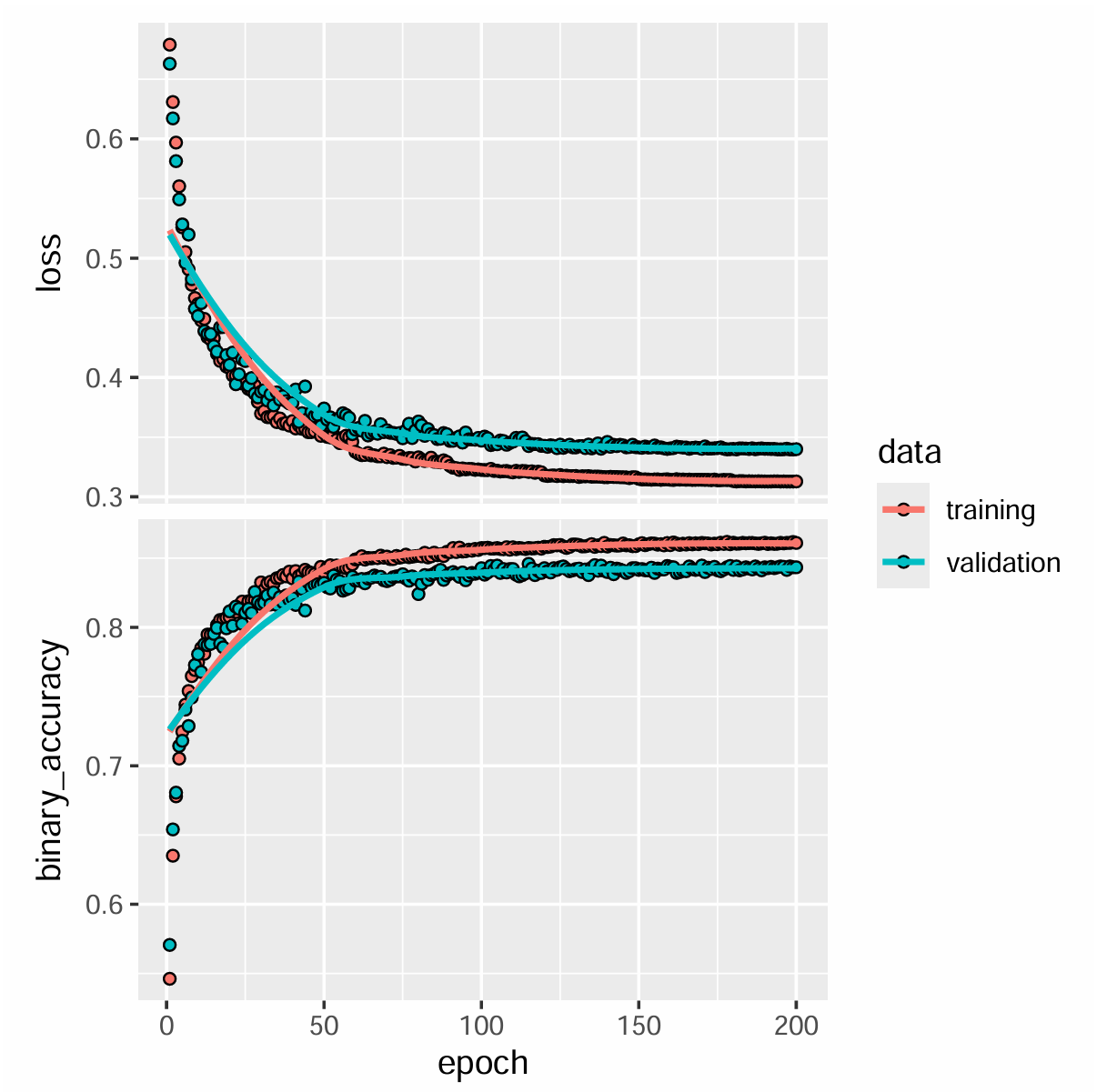}
	\caption{Performance neural network classifier over epochs in training and test data. Top row: training and validation binary cross entropy loss at the end of every epoch. Bottom row: training and validation accuracy  at the end of every epoch.}
	\label{fig:chap4NN1}
\end{figure}
\clearpage

\subsection{Simplified vine copula fit to 5-dimensional illustration data:}
\begin{table}[h]
	\centering
    \resizebox{\textwidth}{!}{%
\begin{tabular}{rrlllrlrr}
	\hline
	tree & edge & conditioned & conditioning & family & rotation & parameters & df & tau \\
	\hline
	1 & 1 & 3, 1 &  & tawn & 0 & 0.64, 0.91, 3.96 & 3 & 0.49 \\
	1 & 2 & 1, 4 &  & bb8 & 0 & 8.00, 0.52 & 2 & 0.48 \\
	1 & 3 & 2, 5 &  & frank & 0 & -2 & 1 & -0.21 \\
	1 & 4 & 4, 5 &  & tawn & 180 & 0.86, 0.89, 1.58 & 3 & 0.31 \\
	2 & 1 & 3, 4 & 1 & tawn & 270 & 0.30, 1.00, 1.76 & 3 & -0.19 \\
	2 & 2 & 1, 5 & 4 & bb7 & 270 & 1.11, 0.41 & 2 & -0.21 \\
	2 & 3 & 2, 4 & 5 & t & 0 & 0.00, 2.42 & 2 & 0.00 \\
	3 & 1 & 3, 5 & 4, 1 & bb7 & 90 & 1.06, 0.08 & 2 & -0.07 \\
	3 & 2 & 1, 2 & 5, 4 & tawn & 0 & 0.30, 0.87, 2.11 & 3 & 0.20 \\
	4 & 1 & 3, 2 & 5, 4, 1 & tawn & 270 & 0.40, 1.00, 2.04 & 3 & -0.27 \\
	\hline
\end{tabular}
}
\caption{Fitted simplified vine copula to the data of the $d=5$ illustration.}
\label{tab:fittedSimplifiedVineExample}
\end{table}
\section{Simulation study}
\label{sec: Appendix B}
\subsection{Simulation setup in the 5-dimensions example}
\begin{table}[h]
		\centering
		\begin{tabular}{|c|c|c|c|}
			\hline
			Tree & Copula & Family & Kendall's tau / parameter function \\ \hline
			First tree &$c_{2,5}$ & Frank & $0.2$\\
			\hline
			First tree &$c_{3,4}$ & Gaussian & $0.3$ \\
			\hline
			First tree &$c_{2,3}$ & Gaussian & $0.4$\\ 
			\hline
			First tree &$c_{1,2}$ & Frank & $0.1$\\
			\hline
			Second tree &$c_{3,5;2}$ & Frank & par\_fun($\mathbf{a}_{3,5;2}$, Frank)($u_2$) \\
			\hline
			Second tree &$c_{2,4;3}$ & Gaussian & par\_fun($\mathbf{a}_{2,4;3}$, Gauss)($u_3$)\\
			\hline
			Second tree &$c_{1,3;2}$ & Gaussian & par\_fun($\mathbf{a}_{1,3;2}$, Gauss)($u_2$)\\
			\hline
			Third tree &$c_{1,5;2,3}$ & Gaussian & par\_fun($\mathbf{a}_{1,5;2,3}$, Gauss)($u_2,u_3$)\\
			\hline
			Third tree &$c_{1,4;2,3}$ & Frank & par\_fun($\mathbf{a}_{1,4;2,3}$, Frank)($u_2,u_3$)\\
			\hline
			Fourth tree &$c_{4,5;1,2,3}$ & Gaussian & par\_fun($\mathbf{a}_{4,5;1,2,3}$, Gauss)($u_1,u_2,u_3$)\\
			\hline
		\end{tabular}
		\caption{Configuration of the 5-dimensional regular vine structure. The varying pair copula parameter functions par\_fun  allowing for linear, quadratic and cubic variation
        are visualized in Figures \ref{fig:plotParamFuncs1d}, \ref{fig:plotParamFuncs2d} and \ref{fig:plotParamFuncs3d} for one, two and three conditioning values, respectively.
        }
		\label{tab:simStudy5dSetup}
	\end{table}
\begin{figure}[h]
	\centering
	\includegraphics[width=0.7\textwidth ]{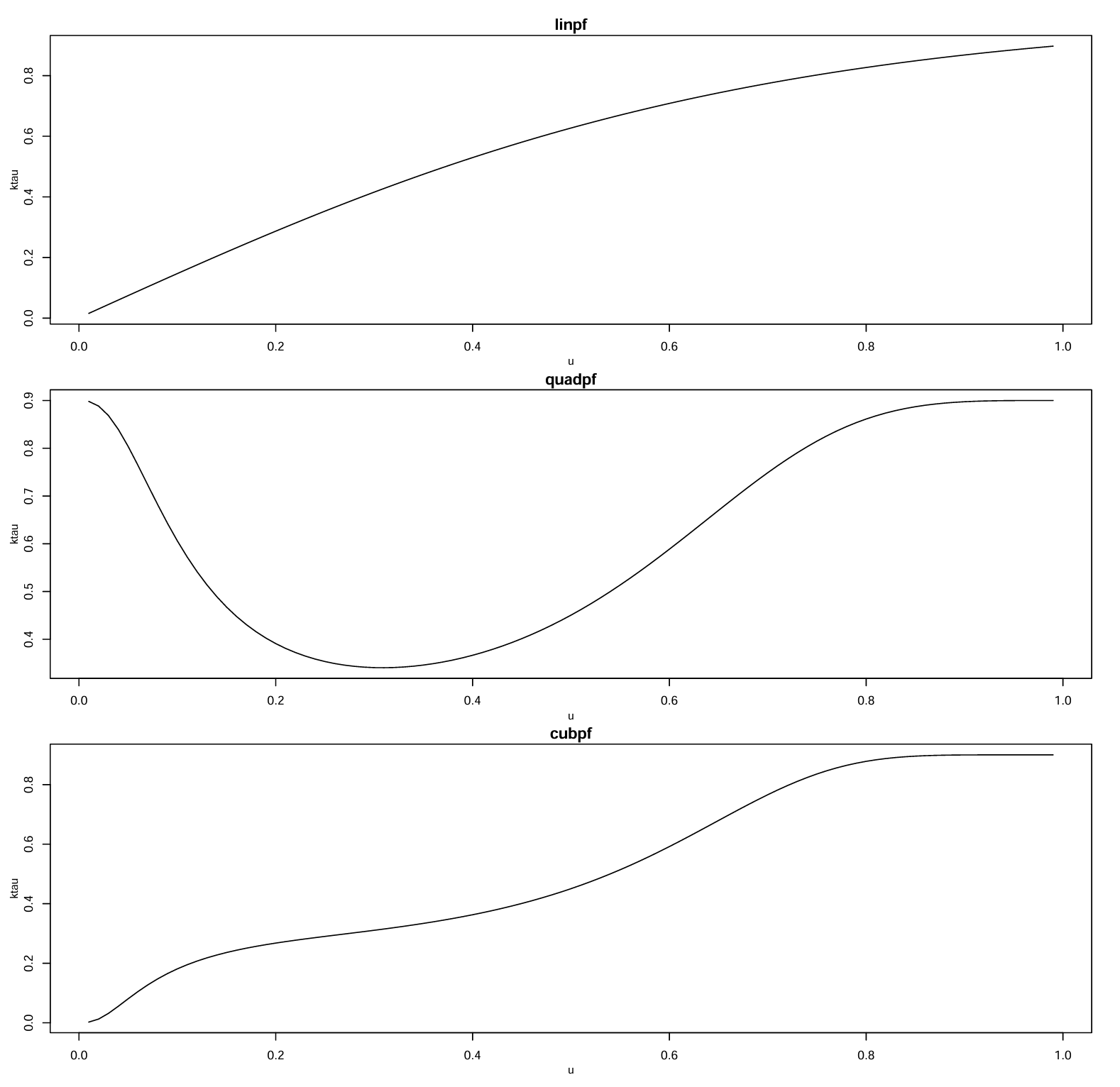}
	\caption{A plot of the conditional linear  (\textbf{linpf}), quadratic (\textbf{quadpf}) and cubic (\textbf{cubpf}) parameter functions for one  conditioning value. 
    }
	\label{fig:plotParamFuncs1d}
\end{figure}

\begin{figure}[h]
	\centering
	\includegraphics[width=0.9\textwidth ]{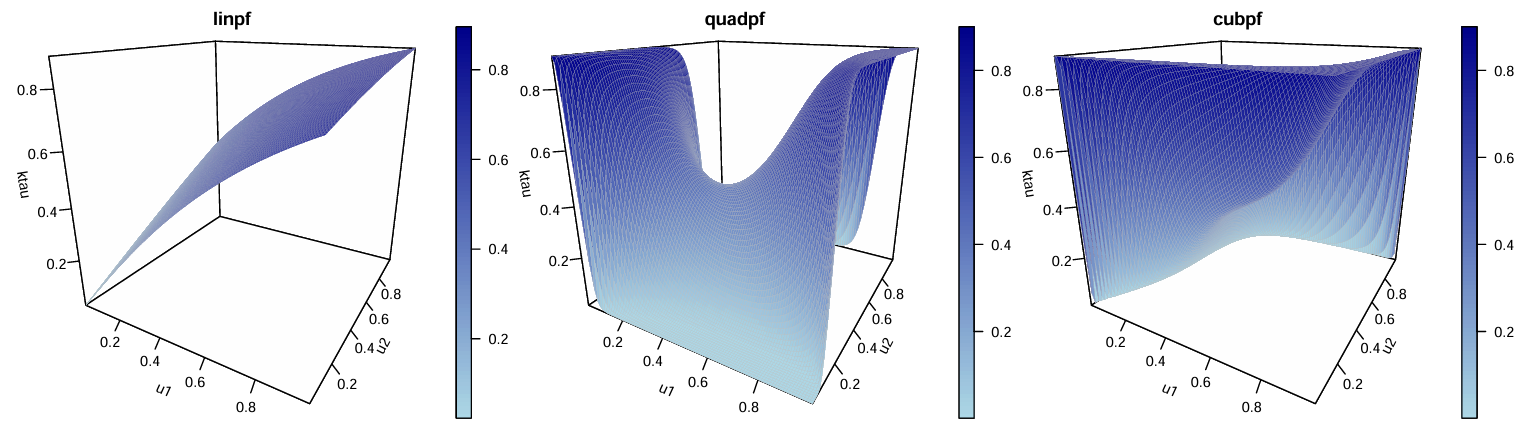}
	\caption{A plot of the conditional linear  (\textbf{linpf}), quadratic (\textbf{quadpf}) and cubic (\textbf{cubpf}) parameter functions for two  conditioning values. 
    }
	\label{fig:plotParamFuncs2d}
\end{figure}

\begin{figure}[h]
	\centering
	\includegraphics[width=\textwidth ]{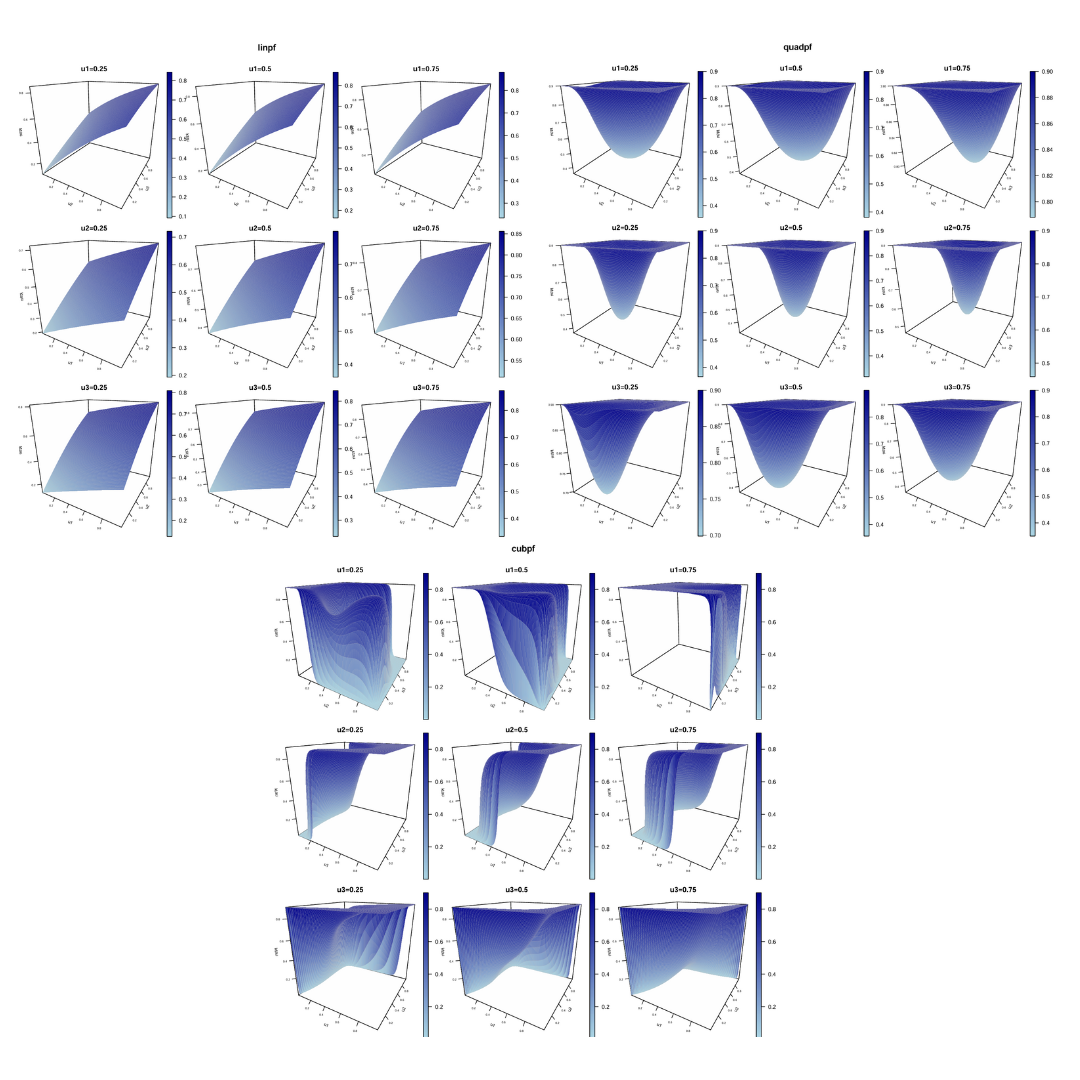}
	\caption{A plot of the conditional linear  (\textbf{linpf}), quadratic (\textbf{quadpf}) and cubic (\textbf{cubpf}) parameter functions for three conditioning values.
    }
	\label{fig:plotParamFuncs3d}
\end{figure}
\clearpage

\subsection{Jensen-Shannon divergence for each configuration between true and noise distribution in the 5-dimensional illustration}

\begin{table}[h]
    \centering
    \begin{tabular}[t]{cccc}
        \hline
        \multicolumn{2}{c}{} & \multicolumn{2}{c}{JSD} \\
        \cmidrule(l{3pt}r{3pt}){3-4}
        $\tau_{max}$ & Par.-fun. & Avg & SE \\
        \hline

        0.3 & constpf & 0.0000 & 0.0000 \\
        0.3 & linpf   & 0.1139 & 0.0058 \\
        0.3 & quadpf  & 0.1651 & 0.0037 \\
        0.3 & cubpf   & 0.1569 & 0.0054 \\

        0.6 & constpf & 0.0000 & 0.0000 \\
        0.6 & linpf   & 0.3058 & 0.0041 \\
        0.6 & quadpf  & 0.4297 & 0.0024 \\
        0.6 & cubpf   & 0.3967 & 0.0030 \\

        0.9 & constpf & 0.0000 & 0.0000 \\
        0.9 & linpf   & 0.5421 & 0.0572 \\
        0.9 & quadpf  & 0.6706 & 0.0243 \\
        0.9 & cubpf   & 0.6607 & 0.0026 \\
        \hline
    \end{tabular}
    \caption{Jensen--Shannon divergence (JSD, \cite{lin2002divergence}) between each configuration 
    and the corresponding configuration with the same $\tau_{\max}$ but with a constant parameter function 
    (simplified vine copula). Based on 1,000 samples and 10 replications.}
    \label{tab:simStudyDistributionDistances}
\end{table}
\clearpage
\subsection{AIC/BIC values for true and approximating vine copula model in the 5-dimensional illustration}
\begin{table}[h]
    \centering
    \resizebox{0.92\textwidth}{!}{%
    \begin{tabular}[t]{ccccccccccc}
        \toprule
        \multicolumn{3}{c}{} 
        & \multicolumn{2}{c}{AIC $c_{noise}$} 
        & \multicolumn{2}{c}{AIC $c_{model}$} 
        & \multicolumn{2}{c}{BIC $c_{noise}$} 
        & \multicolumn{2}{c}{BIC $c_{model}$} \\
        \cmidrule(l{3pt}r{3pt}){4-5}
        \cmidrule(l{3pt}r{3pt}){6-7}
        \cmidrule(l{3pt}r{3pt}){8-9}
        \cmidrule(l{3pt}r{3pt}){10-11}
        $n_{true}$ & $\tau_{max}$ & Par.-fun. 
        & Avg. & SE & Avg. & SE & Avg. & SE & Avg. & SE \\
        \midrule

        1000 & 0.3 & constpf & -1061.2 & 30.5 & -397.5 & 27.9 & -1008.2 & 28.2 & 1329.0 & 28.2 \\
        1000 & 0.3 & linpf   & -1120.3 & 22.6 & -523.5 & 32.8 & -1058.4 & 32.7 & 1211.8 & 32.7 \\
        1000 & 0.3 & quadpf  & -1386.5 & 33.7 & -754.7 & 32.1 & -1328.1 & 32.6 &  977.2 & 32.6 \\
        1000 & 0.3 & cubpf   & -1256.4 & 13.9 & -665.3 & 29.7 & -1190.7 & 28.6 & 1074.0 & 28.6 \\
        \addlinespace

        1000 & 0.6 & constpf & -2061.2 & 36.4 & -1412.6 & 41.6 & -2001.8 & 41.2 &  320.3 & 41.2 \\
        1000 & 0.6 & linpf   & -2250.8 & 49.9 & -1772.2 & 51.7 & -2005.2 & 68.9 &  147.0 & 68.9 \\
        1000 & 0.6 & quadpf  & -3260.5 & 24.2 & -2763.9 & 38.8 & -3057.9 & 64.0 & -887.7 & 64.0 \\
        1000 & 0.6 & cubpf   & -2632.5 & 38.1 & -2279.1 & 47.9 & -2372.2 & 82.3 & -345.2 & 82.3 \\
        \addlinespace

        1000 & 0.9 & constpf & -3737.9 & 25.4 & -3087.3 & 28.6 & -3674.1 & 27.2 & -1350.0 & 27.2 \\
        1000 & 0.9 & linpf   & -5909.4 & 35.7 & -5445.9 & 40.5 & -4373.8 & 98.0 & -2236.8 & 98.0 \\
        1000 & 0.9 & quadpf  & -5664.1 & 72.9 & -5416.2 & 103.9 & -4136.9 & 119.0 & -2215.5 & 119.0 \\
        1000 & 0.9 & cubpf   & -5009.8 & 65.6 & -4801.3 & 70.2 & -3852.8 & 83.7 & -1970.8 & 83.7 \\
        \addlinespace

        10000 & 0.3 & constpf & -10738.8 & 63.8 & -10173.0 & 114.1 & -10658.1 & 114.5 & -7633.5 & 114.5 \\
        10000 & 0.3 & linpf   & -11066.8 & 58.4 & -11233.8 & 62.6 & -10855.9 & 129.9 & -8564.3 & 129.9 \\
        10000 & 0.3 & quadpf  & -14227.3 & 110.6 & -14307.1 & 181.4 & -13534.3 & 272.8 & -11155.4 & 272.8 \\
        10000 & 0.3 & cubpf   & -12605.5 & 137.9 & -12772.9 & 169.3 & -11408.4 & 430.3 & -9117.1 & 430.3 \\
        \addlinespace

        10000 & 0.6 & constpf & -20585.1 & 100.8 & -20047.5 & 105.0 & -20493.5 & 105.7 & -17497.2 & 105.7 \\
        10000 & 0.6 & linpf   & -23327.8 & 152.2 & -25545.4 & 114.9 & -17458.2 & 192.6 & -17217.0 & 192.6 \\
        10000 & 0.6 & quadpf  & -33124.8 & 133.7 & -36857.5 & 255.4 & -25655.2 & 360.2 & -26929.1 & 360.2 \\
        10000 & 0.6 & cubpf   & -27462.7 & 83.7 & -31620.0 & 103.5 & -20431.5 & 297.2 & -22130.1 & 297.2 \\
        \addlinespace

        10000 & 0.9 & constpf & -37388.5 & 64.8 & -37427.5 & 117.5 & -37283.9 & 118.0 & -34864.2 & 118.0 \\
        10000 & 0.9 & linpf   & -64266.0 & 139.0 & -70244.3 & 331.3 & -55552.0 & 370.6 & -59071.6 & 370.6 \\
        10000 & 0.9 & quadpf  & -59646.7 & 137.3 & -74861.0 & 594.2 & -50800.3 & 598.3 & -63555.9 & 598.3 \\
        10000 & 0.9 & cubpf   & -56038.3 & 240.9 & -67520.8 & 556.4 & -46875.7 & 557.9 & -55899.5 & 557.9 \\
        \bottomrule
    \end{tabular}
    }
    \caption{AIC and BIC values for 
$c_{noise}$ (simplified vine copula with all families allowed) 
and for $c_{true}$ across all scenarios. 
Avg denotes the average criterion value and SE the empirical standard error, 
both computed over 10 replications.}
   
    \label{tab:all_noise_simulation_study_AIC_BIC}
\end{table}
\clearpage
\section{Applications}
\label{sec: Appendix C}
\subsection{Abalone data}
\begin{table}[h]
	\centering
    \resizebox{\textwidth}{!}{
	\begin{tabular}{rrlllrlrrr}
		\toprule
		tree & edge & conditioned & conditioning & family & rotation & parameters & df & tau & loglik\\
		\midrule
		1 & 1 & 1, 2 &  & bb6 & 180 & 1.86, 6.44 & 2.0 & 0.89 & 7728.0\\
		1 & 2 & 2, 4 &  & tawn & 180 & 0.95, 1.00, 6.81 & 3.0 & 0.82 & 6466.6\\
		1 & 3 & 6, 4 &  & bb6 & 180 & 1.4, 6.0 & 2.0 & 0.86 & 6733.8\\
		1 & 4 & 5, 4 &  & bb6 & 180 & 2.0, 5.3 & 2.0 & 0.88 & 7253.1\\
		1 & 5 & 3, 7 &  & tawn & 180 & 0.99, 1.00, 4.10 & 3.0 & 0.75 & 4292.8\\
		1 & 6 & 4, 7 &  & bb6 & 180 & 2.73, 3.66 & 2.0 & 0.86 & 6715.4\\
		2 & 1 & 1, 4 & 2 & t & 0 & 0.41, 12.69 & 2.0 & 0.27 & 348.4\\
		2 & 2 & 2, 7 & 4 & tll & 0 &  & 89.9 & 0.19 & 311.6\\
		2 & 3 & 6, 5 & 4 & tll & 0 &  & 112.0 & -0.15 & 306.7\\
		2 & 4 & 5, 7 & 4 & tll & 0 &  & 117.8 & -0.46 & 1406.2\\
		2 & 5 & 3, 4 & 7 & t & 0 & 0.21, 18.58 & 2.0 & 0.14 & 106.4\\
		3 & 1 & 1, 7 & 4, 2 & bb8 & 90 & 1.15, 0.96 & 2.0 & -0.06 & 30.0\\
		3 & 2 & 2, 5 & 7, 4 & bb8 & 180 & 1.26, 0.90 & 2.0 & 0.08 & 31.7\\
		3 & 3 & 6, 7 & 5, 4 & tll & 0 &  & 115.5 & -0.31 & 673.2\\
		3 & 4 & 5, 3 & 7, 4 & bb8 & 90 & 1.31, 0.88 & 2.0 & -0.09 & 45.5\\
		4 & 1 & 1, 5 & 7, 4, 2 & gaussian & 0 & 0.08 & 1.0 & 0.05 & 13.6\\
		4 & 2 & 2, 6 & 5, 7, 4 & frank & 0 & 0.3 & 1.0 & 0.03 & 4.1\\
		4 & 3 & 6, 3 & 7, 5, 4 & bb8 & 180 & 1.11, 0.97 & 2.0 & 0.05 & 14.1\\
		5 & 1 & 1, 6 & 5, 7, 4, 2 & frank & 0 & 0.6 & 1.0 & 0.07 & 20.6\\
		5 & 2 & 2, 3 & 6, 5, 7, 4 & joe & 180 & 1.08 & 1.0 & 0.04 & 35.5\\
		6 & 1 & 1, 3 & 6, 5, 7, 4, 2 & frank & 0 & -0.39 & 1.0 & -0.04 & 8.6\\
		\bottomrule
	\end{tabular}
    }
	\caption{Fitted simplified vine copula to the Abalone data. 
    }
	\label{tab:fittedSimplifiedVineAbaloneExample}
\end{table}

\begin{figure}[h!]
	\centering
	\includegraphics[width=0.4\textwidth ]{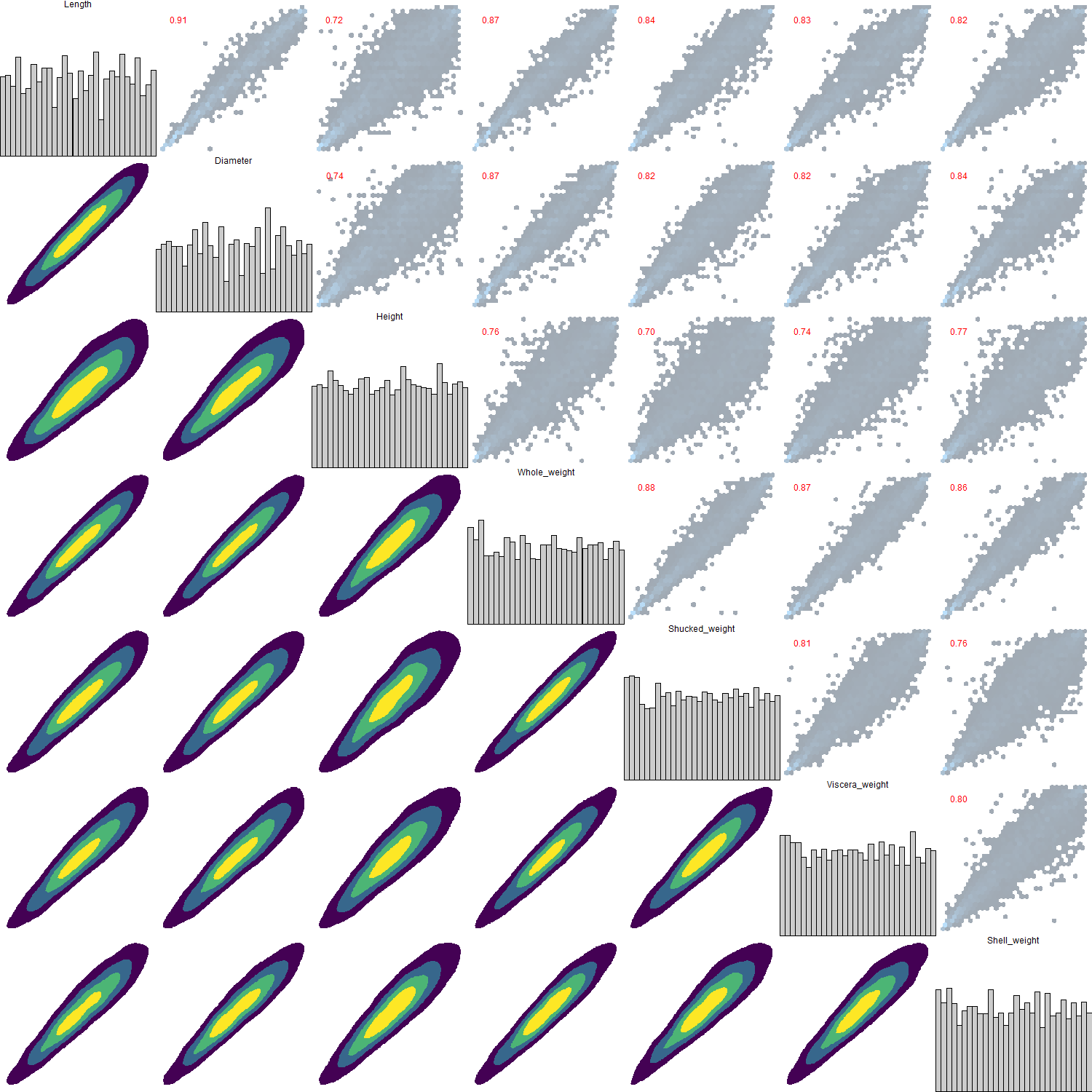}
    \hspace*{.5cm}
    \includegraphics[width=0.4\textwidth ]{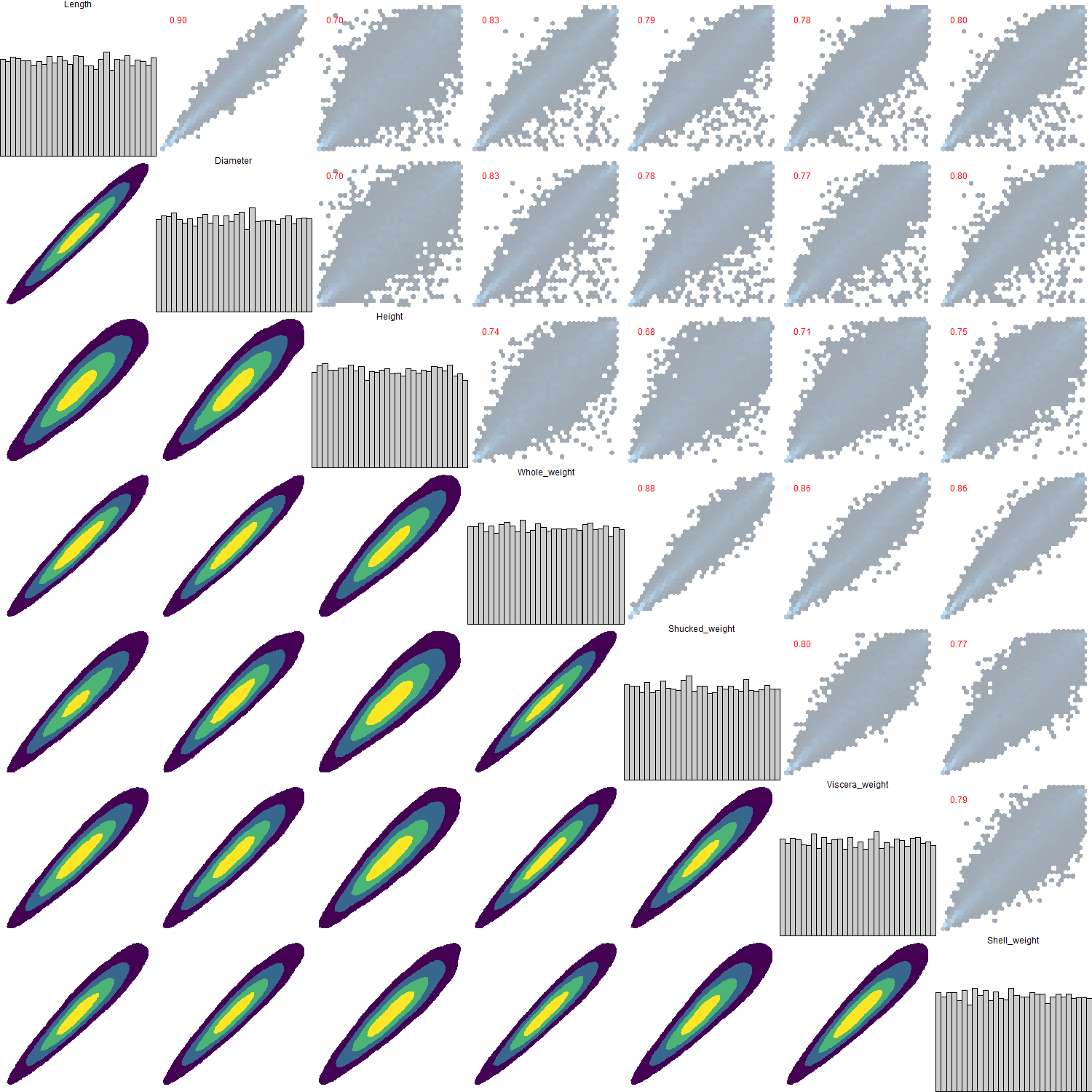}
	\caption{Left: Abalone pseudo-copula data (upper triangle), with marginal $N(0,1)$ contours (lower) and histograms (diagonal). 
Right: 10{,}000 samples from the 20{,}885 fitted via the simplified vine copula model.
    }
	\label{fig:PairPlotOrigAbaloneExample}
\end{figure}
\clearpage

\subsection{Magic Gamma Telescope data set }

\begin{figure}[h]
	\centering
	\includegraphics[width=0.45\textwidth ]{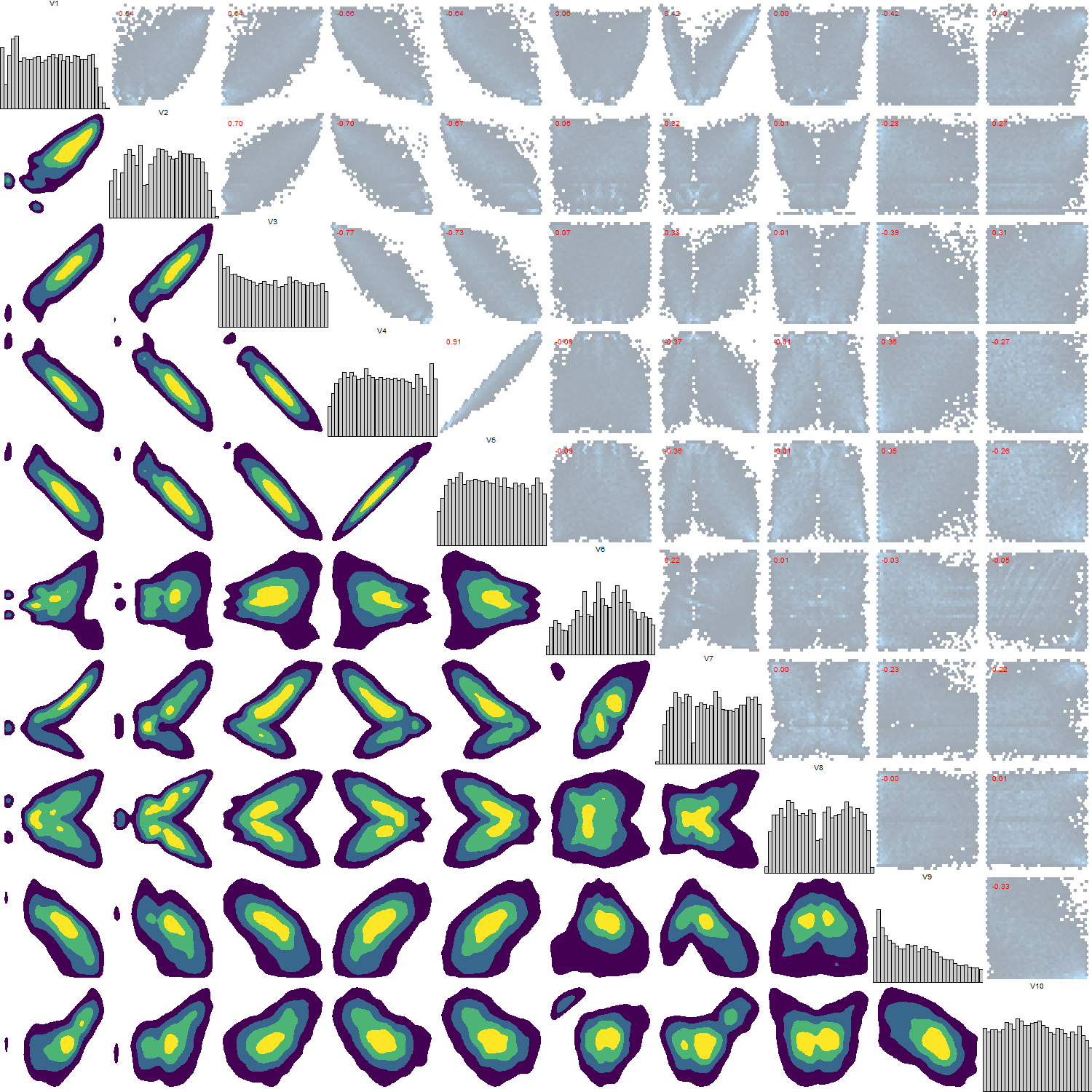}
    \hspace*{.5cm}
    \includegraphics[width=0.45\textwidth ]{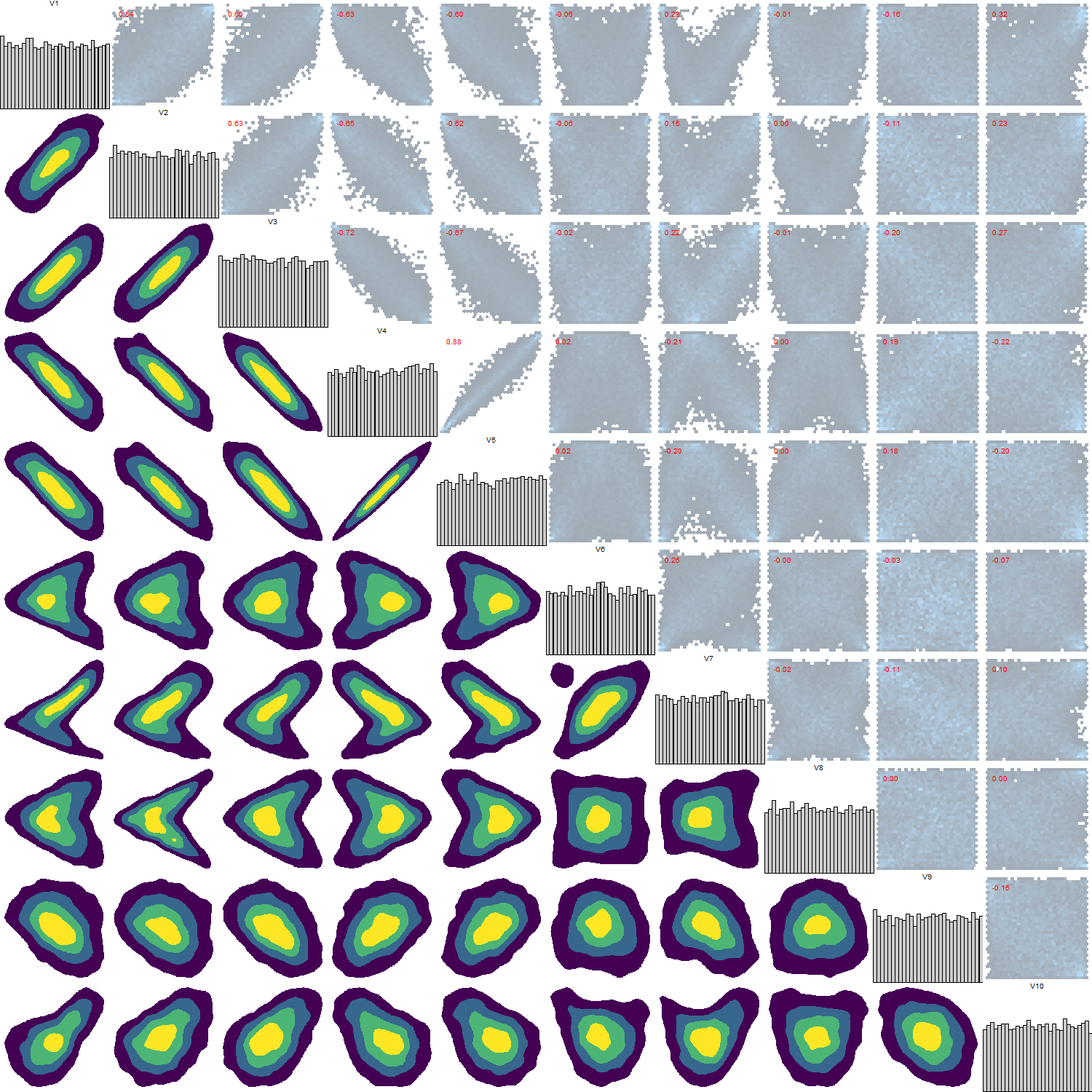}
	\caption{Left: Magic Gamma Telescope pseudo-copula data (upper triangle), with marginal $N(0,1)$ contours (lower) and histograms (diagonal). 
Right: 10{,}000 samples from the 38{,}040 fitted via the simplified vine copula model.
    }
	\label{fig:PairPlotOrigMagicExample}
\end{figure}

\begin{table}[h]
	\centering
    \resizebox{\textwidth}{!}{
	\begin{tabular}{rrlllrlrrr}
		\toprule
		tree & edge & conditioned & conditioning & family & rotation & parameters & df & tau & loglik\\
		\midrule
		1 & 1 & 8, 2 &  & tll & 0 &   & 357.3 & 0.00 & 11420.7\\
		1 & 2 & 9, 3 &  & tll & 0 &   & 177.9 & -0.20 & 1346.8\\
		1 & 3 & 2, 4 &  & tll & 0 &   & 215.5 & -0.65 & 13014.7\\
		1 & 4 & 6, 7 &  & tll & 0 &   & 164.6 & 0.25 & 7088.3\\
		1 & 5 & 5, 4 &  & gumbel & 180 & 8.71 & 1.0 & 0.89 & 33142.2\\
		1 & 6 & 7, 1 &  & tll & 0 &   & 157.9 & 0.24 & 15636.8\\
		1 & 7 & 3, 4 &  & tll & 0 &   & 186.7 & -0.72 & 17431.7\\
		1 & 8 & 4, 1 &  & tll & 0 &   & 201.4 & -0.63 & 12269.9\\
		1 & 9 & 1, 10 &  & tll & 0 &   & 178.4 & 0.33 & 4198.8\\
		\addlinespace
		2 & 1 & 8, 4 & 2 & tll & 0 &   & 177.4 & 0.00 & 417.6\\
		2 & 2 & 9, 4 & 3 & tll & 0 &   & 145.0 & 0.03 & 182.3\\
		2 & 3 & 2, 3 & 4 & bb7 & 180 & 1.03, 0.41 & 2.0 & 0.18 & 886.5\\
		2 & 4 & 6, 1 & 7 & tll & 0 &   & 111.9 & -0.08 & 2844.2\\
		2 & 5 & 5, 3 & 4 & tawn & 180 & 0.30, 0.75, 1.50 & 3.0 & 0.13 & 347.3\\
		2 & 6 & 7, 4 & 1 & bb7 & 270 & 1.01, 0.09 & 2.0 & -0.05 & 58.0\\
		2 & 7 & 3, 1 & 4 & tll & 0 &   & 135.8 & 0.17 & 1012.9\\
		2 & 8 & 4, 10 & 1 & tll & 0 &   & 155.4 & 0.08 & 411.1\\
		\addlinespace
		3 & 1 & 8, 3 & 4, 2 & joe & 90 & 1.02 & 1.0 & -0.01 & 5.5\\
		3 & 2 & 9, 2 & 4, 3 & tll & 0 &   & 167.2 & 0.13 & 981.9\\
		3 & 3 & 2, 5 & 3, 4 & t & 0 & 0.06, 14.45 & 2.0 & 0.04 & 94.8\\
		3 & 4 & 6, 4 & 1, 7 & clayton & 90 & 0.07 & 1.0 & -0.03 & 23.5\\
		3 & 5 & 5, 1 & 3, 4 & bb8 & 0 & 1.16, 0.93 & 2.0 & 0.05 & 97.6\\
		3 & 6 & 7, 3 & 4, 1 & t & 0 & 0.13, 12.56 & 2.0 & 0.08 & 130.9\\
		3 & 7 & 3, 10 & 1, 4 & tawn & 0 & 0.41, 0.43, 1.50 & 3.0 & 0.13 & 409.5\\
		\addlinespace
		4 & 1 & 8, 9 & 3, 4, 2 & t & 0 & 0.00, 34.02 & 2.0 & 0.00 & 4.0\\
		4 & 2 & 9, 5 & 2, 4, 3 & clayton & 180 & 0.04 & 1.0 & 0.02 & 22.9\\
		4 & 3 & 2, 1 & 5, 3, 4 & tll & 0 &   & 199.9 & -0.05 & 2260.9\\
		4 & 4 & 6, 3 & 4, 1, 7 & tll & 0 &   & 127.9 & -0.01 & 235.8\\
		4 & 5 & 5, 10 & 1, 3, 4 & clayton & 0 & 0.03 & 1.0 & 0.01 & 9.5\\
		4 & 6 & 7, 10 & 3, 4, 1 & tll & 0 &   & 141.7 & -0.03 & 227.3\\
		\addlinespace
		5 & 1 & 8, 5 & 9, 3, 4, 2 & t & 0 & 0.00, 21.99 & 2.0 & 0.00 & 18.5\\
		5 & 2 & 9, 1 & 5, 2, 4, 3 & tll & 0 &   & 142.2 & -0.03 & 167.4\\
		5 & 3 & 2, 10 & 1, 5, 3, 4 & tll & 0 &   & 162.6 & 0.06 & 385.7\\
		5 & 4 & 6, 10 & 3, 4, 1, 7 & bb8 & 90 & 1.27, 0.77 & 2.0 & -0.05 & 54.1\\
		5 & 5 & 5, 7 & 10, 1, 3, 4 & t & 0 & -0.01, 36.42 & 2.0 & 0.00 & 9.8\\
		\addlinespace
		6 & 1 & 8, 1 & 5, 9, 3, 4, 2 & tll & 0 &   & 140.0 & -0.01 & 245.2\\
		6 & 2 & 9, 10 & 1, 5, 2, 4, 3 & tll & 0 &   & 161.7 & -0.11 & 585.5\\
		6 & 3 & 2, 7 & 10, 1, 5, 3, 4 & tll & 0 &   & 135.7 & -0.04 & 359.6\\
		6 & 4 & 6, 5 & 10, 3, 4, 1, 7 & bb8 & 270 & 1.08, 0.97 & 2.0 & -0.04 & 29.5\\
		\addlinespace
		7 & 1 & 8, 10 & 1, 5, 9, 3, 4, 2 & indep & 0 &  & 0.0 & 0.00 & 0.0\\
		7 & 2 & 9, 7 & 10, 1, 5, 2, 4, 3 & clayton & 90 & 0.07 & 1.0 & -0.03 & 37.4\\
		7 & 3 & 2, 6 & 7, 10, 1, 5, 3, 4 & t & 0 & -0.07, 10.20 & 2.0 & -0.05 & 78.7\\
		\addlinespace
		8 & 1 & 8, 7 & 10, 1, 5, 9, 3, 4, 2 & joe & 90 & 1.03 & 1.0 & -0.01 & 6.1\\
		8 & 2 & 9, 6 & 7, 10, 1, 5, 2, 4, 3 & frank & 0 & -0.23 & 1.0 & -0.03 & 12.9\\
		\addlinespace
		9 & 1 & 8, 6 & 7, 10, 1, 5, 9, 3, 4, 2 & t & 0 & 0.00, 8.49 & 2.0 & 0.00 & 41.9\\
		\bottomrule
	\end{tabular}
    }
	\caption{Fitted simplified vine copula to the Magic Gamma Telescope data.  
    }
	\label{tab:fittedSimplifiedVineMagicExample}
\end{table}

\clearpage

\end{appendices}


\bibliography{sn-bibliography}

\end{document}